\pdfoutput=1
\RequirePackage{fix-cm}
\documentclass[twocolumn,epjc3]{svjour3}
\smartqed
\usepackage{graphicx}
\usepackage{amsmath,amssymb}
\usepackage{url}
\usepackage[final]{microtype}
\graphicspath{{figures/}{paper/figures/}}

\newcommand{\dminLHClo}{2.9}
\newcommand{\dminLHChi}{8.2}
\newcommand{\dminRHIClo}{5.9}
\newcommand{\dminRHIChi}{12.5}
\newcommand{\wMaxA}{0.75}
\newcommand{\dminAlo}{11}
\newcommand{\dminAhi}{21}
\newcommand{\dmaxInj}{0.25}
\newcommand{\dmaxInjPct}{25}
\newcommand{\dminGridLo}{0.01}
\newcommand{\nDepths}{10}
\newcommand{\nWidths}{7}
\newcommand{\nWidthsWord}{seven}
\newcommand{\widthList}{0.3, 0.5, 0.75, 1, 1.5, 2, 3}
\newcommand{\dminWtwoLo}{18}
\newcommand{\dminWtwoHi}{24}
\newcommand{\dminWtwoE}{200~GeV, 2.76~TeV and 5.36~TeV}
\newcommand{\pMinAll}{0.23}
\newcommand{\nConfAll}{546}
\newcommand{\nToys}{200}
\newcommand{\pRes}{0.005}
\newcommand{\binomHalf}{0.035}
\newcommand{\pNomBothMin}{0.065}
\newcommand{\qDgMax}{13.4}

\newcommand{\qNinetyFiveLo}{3.0}
\newcommand{\qNinetyFiveHi}{4.4}
\newcommand{\chiNdfNomRHIClo}{0.05}
\newcommand{\chiNdfNomRHIChi}{0.31}
\newcommand{\chiNdfScanRHIChi}{1.2}
\newcommand{\chiNdfNomLHClo}{0.04}
\newcommand{\chiNdfNomLHChi}{0.35}
\newcommand{\qEdgeMax}{0.35}
\newcommand{\qEdgeMaxAll}{1.30}
\newcommand{\chiBoxLo}{7.6}
\newcommand{\chiBoxHi}{14.5}
\newcommand{\chiBoxELo}{3.0}
\newcommand{\chiBoxEHi}{4.2}
\newcommand{\aliceRatioLoE}{0.99}
\newcommand{\aliceRatioHiE}{1.05}
\newcommand{\aliceRatioLoF}{0.95}
\newcommand{\aliceRatioHiF}{1.04}
\newcommand{\kappaNom}{1.19}
\newcommand{\kappaErr}{0.09}
\newcommand{\kappaSmooth}{1.22}
\newcommand{\sigScale}{5.3}
\newcommand{\kappaFlat}{0.85}
\newcommand{\kappaFlatErr}{0.09}
\newcommand{\fEA}{1.25}
\newcommand{\fEAPull}{+0.7}
\newcommand{\fEAflat}{0.85}
\newcommand{\fEAflatPull}{-0.1}
\newcommand{\fEB}{1.26}
\newcommand{\fEBPull}{+0.9}
\newcommand{\fEBflat}{0.91}
\newcommand{\fEBflatPull}{+0.6}
\newcommand{\kappaSpecies}{0.96}
\newcommand{\kappaSpeciesErr}{0.10}
\newcommand{\netbTerm}{0.22}
\newcommand{\fwdShareLo}{47}
\newcommand{\fwdShareHi}{52}
\newcommand{\kernPfwd}{10.0}
\newcommand{\netpFwd}{29.4}
\newcommand{\kappaStar}{1.28}
\newcommand{\kappaStarErr}{0.15}
\newcommand{\kStarFactor}{1.08}
\newcommand{\kTransfer}{1.06}
\newcommand{\kTransferE}{1.17}

\newcommand{\sLoE}{1.09}
\newcommand{\sHiE}{1.09}
\newcommand{\dOneLoE}{1.3}
\newcommand{\dOneHiE}{1.3}
\newcommand{\pullLoE}{2.5}
\newcommand{\pullHiE}{2.5}
\newcommand{\sKLoE}{1.09}
\newcommand{\sKHiE}{1.10}

\newcommand{\dOneKHiE}{1.6}

\newcommand{\sELoE}{1.10}
\newcommand{\sEHiE}{1.12}
\newcommand{\dOneEHiE}{2.3}
\newcommand{\sNFLoE}{0.94}
\newcommand{\sNFHiE}{1.29}

\newcommand{\dOneVarHiE}{1.4}

\newcommand{\sLoF}{1.00}
\newcommand{\sHiF}{1.03}

\newcommand{\pullLoF}{2.6}
\newcommand{\pullHiF}{2.9}
\newcommand{\sKLoF}{1.03}
\newcommand{\sKHiF}{1.06}

\newcommand{\dOneKHiF}{0.4}

\newcommand{\sELoF}{1.08}
\newcommand{\sEHiF}{1.11}
\newcommand{\dOneEHiF}{1.3}
\newcommand{\sNFLoF}{0.83}
\newcommand{\sNFHiF}{1.27}

\newcommand{\sLoG}{0.84}
\newcommand{\sHiG}{0.92}
\newcommand{\dOneLoG}{0.8}
\newcommand{\dOneHiG}{2.6}
\newcommand{\pullLoG}{1.2}
\newcommand{\pullHiG}{1.7}
\newcommand{\sKLoG}{0.88}
\newcommand{\sKHiG}{0.95}

\newcommand{\dOneKHiG}{1.4}

\newcommand{\sELoG}{0.96}
\newcommand{\sEHiG}{1.02}
\newcommand{\dOneEHiG}{0.1}

\newcommand{\sNFG}{\mbox{0.68--1.12}}
\newcommand{\dOneVarHiG}{2.6}
\newcommand{\varMaxRuleG}{flat}
\newcommand{\varMaxFamG}{box+edge}
\newcommand{\pullKLo}{2.0}
\newcommand{\pullKHi}{2.5}
\newcommand{\pullELo}{1.1}
\newcommand{\pullEHi}{1.9}
\newcommand{\pullLoEF}{2.5}
\newcommand{\pullHiEF}{2.9}

\newcommand{\yeNearLo}{7.4}
\newcommand{\yeNearHi}{8.7}
\newcommand{\aeNearLo}{0.28}
\newcommand{\aeNearHi}{0.81}
\newcommand{\yeBroadLo}{4.5}
\newcommand{\yeBroadHi}{4.5}
\newcommand{\aeBroadLo}{1.38}
\newcommand{\aeBroadHi}{1.39}
\newcommand{\edgeBroadList}{box+edge and gG+edge at 5.02~TeV}

\newcommand{\nVariantsWord}{six}
\newcommand{\nVariantsMinusOneWord}{five}

\newcommand{\fineLo}{0.66}
\newcommand{\fineHi}{1.50}
\newcommand{\kapLo}{0.80}
\newcommand{\kapHi}{1.30}
\newcommand{\kLowFactor}{0.94}
\newcommand{\dOneKlowHiG}{4.2}
\newcommand{\dOneKlowRel}{\ge}
\newcommand{\nKlowOut}{at least one}
\newcommand{\coarseStep}{0.1}
\newcommand{\varSpreadText}{less than one grid step (the same grid point for every rule)}
\newcommand{\varEdgeText}{which stays below the 95\% threshold of 3.84 for every rule and family}
\newcommand{\ovEtap}{-2.61}
\newcommand{\ovRatio}{1.09}
\newcommand{\ovErr}{0.09}
\newcommand{\npartPhobos}{344}
\newcommand{\npartPhobosErr}{11}
\newcommand{\hijRcal}{0.99}
\newcommand{\hijRcalA}{1.09}

\newcommand{\hijRlhc}{\mbox{1.04--1.05}}

\newcommand{\hijTransfer}{\mbox{1.05--1.06}}

\newcommand{\hijTransferOwn}{\mbox{1.05--1.07}}

\newcommand{\hijKernelLHCsc}{1.8}
\newcommand{\hijBiasRHICmax}{0.50}
\newcommand{\hijDevRHICmax}{10}
\newcommand{\hijDevInnerLo}{6}
\newcommand{\hijDevInnerHi}{9}

\newcommand{\hijFalseQ}{12.7}
\newcommand{\hijFalseFam}{box}
\newcommand{\hijFalseQnf}{4.4}
\newcommand{\hijExactInnerLo}{5}
\newcommand{\hijExactInnerHi}{9}
\newcommand{\hijExactAllMax}{15}
\newcommand{\hijExactQ}{10.5}
\newcommand{\hijTrueDip}{1.4}
\newcommand{\hijYbmQ}{12.6}
\newcommand{\hijEtaThreeQ}{1.9}
\newcommand{\hijEdgeQ}{12.7}

\newcommand{\hijPtStarD}{2}

\newcommand{\hijPtNomD}{1}

\newcommand{\epoRcal}{1.11}
\newcommand{\epoRcalA}{1.22}

\newcommand{\epoRlhc}{\mbox{1.28--1.31}}

\newcommand{\epoTransfer}{\mbox{1.16--1.18}}

\newcommand{\epoTransferOwn}{1.03}

\newcommand{\epoKernelLHCsc}{1.1}
\newcommand{\epoBiasRHICmax}{0.59}
\newcommand{\epoDevRHICmax}{36}
\newcommand{\epoDevInnerLo}{3}
\newcommand{\epoDevInnerHi}{4}

\newcommand{\epoFalseQ}{5.8}
\newcommand{\epoFalseFam}{dG}
\newcommand{\epoFalseQnf}{4.1}
\newcommand{\epoExactInnerLo}{3}
\newcommand{\epoExactInnerHi}{3}
\newcommand{\epoExactAllMax}{22}
\newcommand{\epoExactQ}{4.1}

\newcommand{\epoYbmQ}{0.4}
\newcommand{\epoEtaThreeQ}{0.4}
\newcommand{\epoEdgeQ}{0.1}

\newcommand{\epoPtStarD}{25}
\newcommand{\epoPtStarKD}{30}
\newcommand{\epoPtNomD}{-2}

\newcommand{\epoNevLow}{300}
\newcommand{\epoNevHigh}{499}
\newcommand{\epoExactMax}{2.8}
\newcommand{\epoEnergyDev}{0.74}
\newcommand{\epoRejected}{1}

\newcommand{\brahmsFoldMid}{0.97}
\newcommand{\brahmsEtaMid}{0.15}
\newcommand{\brahmsFoldMin}{0.66}
\newcommand{\brahmsPhobosLo}{0.88}
\newcommand{\brahmsPhobosHi}{1.10}
\newcommand{\etaMaxE}{5.4}
\newcommand{\etapMinE}{-2.61}
\newcommand{\etaMaxF}{4.9}
\newcommand{\etapMinF}{-3.71}
\newcommand{\etaMaxG}{2.5}
\newcommand{\etapMinG}{-6.15}
\newcommand{\ybeamA}{3.04}
\newcommand{\ybeamB}{4.20}
\newcommand{\ybeamC}{4.93}
\newcommand{\ybeamD}{5.36}

\newcommand{\npartE}{382.8}
\newcommand{\npartF}{389.3}
\newcommand{\npartG}{382.3}
\newcommand{\npartA}{337}

\newcommand{\npartC}{340}
\newcommand{\npartD}{344}
\newcommand{\fOther}{2}
\newcommand{\fOtherLo}{0}
\newcommand{\fOtherHi}{4}
\newcommand{\ptScaleVar}{3.7}
\newcommand{\nCovScan}{15}
\newcommand{\yieldPi}{1475}
\newcommand{\yieldPubPi}{1465}
\newcommand{\yieldK}{218}
\newcommand{\yieldPubK}{218}
\newcommand{\yieldP}{67}
\newcommand{\yieldPubP}{67}

\newcommand{\jacHijAbsLo}{4.4}
\newcommand{\jacHijAbsHi}{4.9}
\newcommand{\jacEpoLo}{-0.5}
\newcommand{\jacEpoHi}{0.3}

\newcommand{\mptPstar}{1.42}
\newcommand{\mptPnom}{1.01}
\newcommand{\mptPlhc}{1.35}
\newcommand{\mptPstarB}{1.30}
\newcommand{\mptPnomB}{0.95}
\newcommand{\mptBiasData}{40}

\newcommand{\pShareVar}{25}
\newcommand{\cB}{2.03}
\newcommand{\Rpi}{1.5}
\newcommand{\Rk}{2}
\newcommand{\nuFlat}{1.5}
\newcommand{\netbDy}{2.04}
\newcommand{\netbE}{27.7--33.2}
\newcommand{\netbDyLo}{1.65}
\newcommand{\netbDyHi}{2.34}
\newcommand{\netbShiftLo}{-0.59}
\newcommand{\netbShiftHi}{+0.41}
\newcommand{\fineStep}{0.02}
\newcommand{\kapStep}{0.05}
\newcommand{\lfLo}{-2.5}
\newcommand{\covNormE}{1}
\newcommand{\covInE}{80}
\newcommand{\covOutE}{77}
\newcommand{\covNormF}{0.6}
\newcommand{\covNormG}{0.1}
\newcommand{\covInF}{80}
\newcommand{\covCmsUnc}{0.37}
\newcommand{\covCmsShape}{0.37}
\newcommand{\covCmsFull}{0.26}
\newcommand{\covCmsEll}{1}
\newcommand{\covVarLo}{50}
\newcommand{\covVarHi}{95}
\newcommand{\covNormVarE}{2}
\newcommand{\covEllLo}{0.5}
\newcommand{\covEllHi}{2}
\newcommand{\hijNevHigh}{2000}
\newcommand{\hijNevLow}{1000}
\newcommand{\hijEnergyDev}{7\times10^{-4}}
\newcommand{\hijExactMax}{0.9}
\newcommand{\nBoot}{100}
\newcommand{\freeZeroLoE}{0.64}
\newcommand{\freeZeroHiE}{0.76}
\newcommand{\freeZeroEdgeE}{0.01}
\newcommand{\freeEnergyLoE}{1.4}
\newcommand{\freeEnergyHiE}{1.5}
\newcommand{\freeEnergyEdgeE}{0.6}

\newcommand{\freeEnergyLoF}{2.4}
\newcommand{\freeEnergyHiF}{2.9}

\newcommand{\systShiftMax}{0.22}

\newcommand{\sSystLoE}{1.00}
\newcommand{\sSystHiE}{1.23}
\newcommand{\sSystLoF}{0.91}
\newcommand{\sSystHiF}{1.23}
\newcommand{\sSystLoG}{0.75}
\newcommand{\sSystHiG}{1.14}
\newcommand{\systOutText}{at 2.76~TeV, with $s>1$, for all three families under strongly correlated PHOBOS uncertainties; at 5.02~TeV, with $s>1$, for one family under strongly correlated PHOBOS uncertainties ($\Delta\chi^2(s{=}1)=4.1$); at 5.36~TeV, with $s<1$, for all three families under uncorrelated PHOBOS uncertainties, for two families under a kernel without rapidity dependence and for one family under the BRAHMS dependence anchored at beam rapidity ($\Delta\chi^2(s{=}1)=4.0$)}
\newcommand{\systUnionText}{at 2.76~TeV ($s>1$) for strongly correlated PHOBOS uncertainties and at 5.36~TeV ($s<1$) for uncorrelated PHOBOS uncertainties}
\newcommand{\systUnionAbs}{2.76~TeV (ratio above) or 5.36~TeV (below)}
\newcommand{\sSystAllLo}{0.75}
\newcommand{\sSystAllHi}{1.23}

\newcommand{\covToys}{100}
\newcommand{\covFracLo}{0.04}
\newcommand{\covFracHi}{0.06}
\newcommand{\covEdgeMax}{27}

\newcommand{\fvKappaErr}{0.10}
\newcommand{\fvSigScale}{7.6}
\newcommand{\fvShiftMax}{0.07}
\newcommand{\fvDOneMax}{1.4}
\newcommand{\fvNFLoE}{0.89}
\newcommand{\fvNFHiE}{1.41}
\newcommand{\fvNFLoF}{0.79}
\newcommand{\fvNFHiF}{1.54}
\newcommand{\fvNFLoG}{0.63}
\newcommand{\fvNFHiG}{1.23}
\newcommand{\fvPullLoEF}{1.6}
\newcommand{\fvPullHiEF}{1.9}
\newcommand{\mirrorPmin}{0.77}
\newcommand{\mirrorDminLo}{6.6}
\newcommand{\mirrorDminHi}{15.3}
\newcommand{\outsideQzeroMax}{0.57}
\newcommand{\outsideDLo}{7.0}
\newcommand{\outsideDHi}{20.0}
\newcommand{\outsideRatioLo}{1.8}
\newcommand{\outsideRatioHi}{2.2}
\newcommand{\clsLoHalf}{4.7}
\newcommand{\clsHiHalf}{9.1}

\newcommand{\clsLoOne}{8.8}
\newcommand{\clsHiOne}{11.5}

\newcommand{\clsAHalf}{18.3}
\newcommand{\lowRateLo}{0.04}
\newcommand{\lowRateHi}{0.18}

\newcommand{\snn}{\sqrt{s_{NN}}}
\newcommand{\dndeta}{dN_{\mathrm{ch}}/d\eta}
\newcommand{\dndy}{dN_{\mathrm{ch}}/dy}
\newcommand{\ybeam}{y_{\mathrm{beam}}}
\newcommand{\Npart}{\langle N_{\mathrm{part}}\rangle}
\newcommand{\rng}[2]{\edef\rngA{#1}\edef\rngB{#2}\ifx\rngA\rngB$#1$\else\mbox{$#1$--$#2$}\fi}

\journalname{Eur. Phys. J. C}

\begin{document}

\title{Midrapidity dip and limiting fragmentation of charged hadrons from RHIC to LHC energies}

\author{Neeraj\thanksref{e1,addr1}
\and Md.~Kaosar Ali Mondal\thanksref{e2,addr1}
\and Amal Sarkar\thanksref{e3,addr1}}

\thankstext{e1}{e-mail: neeraj.neeraj@cern.ch (corresponding author)}
\thankstext{e2}{e-mail: mohammad.mondal@cern.ch}
\thankstext{e3}{e-mail: amal@iitmandi.ac.in}

\institute{School of Physical Sciences, Indian Institute of Technology Mandi, Kamand, Himachal Pradesh 175005, India\label{addr1}}

\date{Received: date / Accepted: date}

\maketitle

\begin{abstract}
The charged-hadron pseudorapidity density $\dndeta$ in central heavy-ion collisions has a
midrapidity depression that deepens with collision energy. Whether it reflects structure in the
rapidity density $\dndy$ or only the kinematics of the $y\to\eta$ transformation has mostly been
argued with effective or model-based Jacobians. We test this without an event generator, with a $y\to\eta$
kernel built from measured identified-hadron spectra. In central Au+Au and Pb+Pb collisions from
$\snn=19.6$~GeV to 5.36~TeV the data do not require a dip in $\dndy$: the decisive null-hypothesis
family gives $p\ge\pMinAll$ in all variants. For dips up to one rapidity unit
wide, the depth giving $p<0.05$ in half the pseudo-experiments is
\rng{\dminLHClo}{\dminLHChi}\% at the LHC and \rng{\dminRHIClo}{\dminRHIChi}\% at 62.4--200~GeV. Generator closure
tests show a false dip at top
RHIC energy caused by the forward region. As the LHC data barely constrain the
fragmentation region, an energy sum rule calibrated on RHIC data, with the RHIC shape of that
region, gives best-fit LHC/RHIC ratios of \rng{\sLoE}{\sHiE},
\rng{\sLoF}{\sHiF} and \rng{\sLoG}{\sHiG} at 2.76, 5.02 and 5.36~TeV for three shape families,
with 95\% intervals over the families of \rng{\sNFLoE}{\sNFHiE}, \rng{\sNFLoF}{\sNFHiF} and \sNFG{}
that include unity, although at 2.76 and 5.02~TeV the fits prefer \rng{\pullLoEF}{\pullHiEF}$\,\sigma$
more energy than calibrated. Over plausible RHIC covariances and kernels the best-fit
ratios span \rng{\sSystAllLo}{\sSystAllHi}, and single variants exclude unity at \systUnionAbs: limiting
fragmentation is neither established nor excluded.
\end{abstract}

\section{Introduction}
\label{sec:intro}

The charged-hadron pseudorapidity density $\dndeta$ is one of the most precisely measured global
observables in heavy-ion collisions. It is available with large acceptance from RHIC to the highest
LHC energies~\cite{Back:2002wb,PHOBOS:2005zhy,PHOBOS:2010eyu,ALICE:2013jfw,ALICE:2016fbt,CMS:2024ykx}.
In central collisions it has a shallow depression at $\eta=0$ that becomes more pronounced with
energy. Pseudorapidity ignores the particle mass: at fixed transverse momentum the Jacobian
$dy/d\eta=p/E$ is smallest at midrapidity, so a rapidity density $\dndy$ that is flat near $y=0$
produces a dip in $\dndeta$ by kinematics alone. How much of the observed dip this accounts for
depends on the transverse-momentum spectra, the particle composition~\cite{Sahoo:2014aca} and the flatness of $\dndy$
near $y=0$.

Earlier studies came to different conclusions, and they used different Jacobians.
In the three-source relativistic diffusion model (RDM), an early LHC prediction with a single
effective mass and transverse momentum found the size of the dip ``essentially determined by the
relative particle content in the three sources, not by the Jacobian''~\cite{Wolschin:2011za}. Later
RDM analyses used an effective Jacobian determined from identified-particle data and found it
``essential to generate the dip''~\cite{Kellers:2019ezz}: near midrapidity ``it generates the
dip''~\cite{Kellers:2020xfa}. Hoelck and
Wolschin, solving a two-dimensional diffusion model for pions, kaons and protons with the full
momentum- and mass-dependent Jacobian, found the dip ``mainly due to the Jacobian
transformation''~\cite{Hoelck:2024cdl}. ALICE converted its 2.76~TeV $\dndeta$ to $\dndy$ with a
Jacobian from its measured midrapidity $\pi$, $K$ and $p$ spectra and found the result consistent with
a flat rapidity plateau of about $\pm1.5$ units around $y=0$, and also with a wide Gaussian of width
3.86~\cite{ALICE:2013jfw}. Event generators provide the full momentum of every particle, but their
spectra need not match the data: HIJING~\cite{Wang:1991hta,Gyulassy:1994ew}, for example, has softer spectra than
measured and therefore a deeper kinematic dip.

The question is addressed here without a model of particle production: do the measured $\dndeta$
require any midrapidity structure in $\dndy$ once the exact $y\to\eta$ kinematics with measured
identified spectra are applied? Candidate $\dndy$ shapes are folded through a kernel built from the
measured $\pi$, $K$ and $p$ spectra and their measured rapidity dependence. The null hypothesis is
a family of smooth dip-free shapes; the alternative adds a dip of free depth and width. The test
statistic is calibrated with pseudo-experiments, its sensitivity is quantified by injecting dips,
and the decision rule was fixed before any $p$-value was computed, although it was not registered
publicly; later changes are listed in \ref{app:amendments}. The analysis is not a replacement for momentum-space models such as
the RDM. It tests what the data require, and a $\dndy$ without a dip in rapidity, such as a
dominant midrapidity source with fragmentation sources, is one of the shapes it allows.

The same machinery bears on a second question. Limiting fragmentation~\cite{Benecke:1969sh}, the
energy independence of particle production near beam rapidity, has been observed at
RHIC~\cite{Back:2002wb}, and ALICE found its 2.76~TeV data consistent with it within the
uncertainties of extrapolating them toward beam rapidity~\cite{ALICE:2013jfw}. LHC data stop
several units before beam rapidity, so statements at the LHC rest on an
extrapolation. Double-Gaussian extrapolations of the ALICE data have been read as violating
limiting fragmentation at 5.02~TeV when the Gaussians are fitted directly in
$\eta$~\cite{Sahoo:2018osl}, and as fulfilling it when they are fitted in rapidity and transformed
with a Jacobian~\cite{Kellers:2019ezz}. In the RDM the fragmentation tail is shaped by the Dirichlet
boundary condition of the numerical solver, and at LHC energies that analysis of limiting
fragmentation ``remains a model-dependent prediction''~\cite{Kellers:2020xfa}. The data alone
are found to leave this extrapolation largely unconstrained, and the constraint provided by an energy
sum rule calibrated on RHIC data is examined.

Section~\ref{sec:data} lists the data and Sect.~\ref{sec:method} describes the kernel and the test.
Section~\ref{sec:results} gives the results on the dip, Sect.~\ref{sec:closure} closure tests with event
generators and Sect.~\ref{sec:brahms} a comparison of identified and inclusive BRAHMS yields. Limiting fragmentation is
discussed in Sect.~\ref{sec:lf}, limitations in Sect.~\ref{sec:discussion}, and
Sect.~\ref{sec:summary} summarizes.

\section{Data}
\label{sec:data}

\paragraph{Pseudorapidity densities.} For Au+Au collisions the data are the PHOBOS 0--6\% central
$\dndeta$ at $\snn=19.6$, 62.4, 130 and 200~GeV from the collaboration's published data
tables~\cite{Back:2002wb,PHOBOS:2005zhy,PHOBOS:2010eyu}. The tables give systematic uncertainties
only, which PHOBOS quotes at 90\% confidence level, with negligible statistical
errors~\cite{Back:2002wb,PHOBOS:2005zhy}. We average the upper and lower uncertainties and divide by
1.645, the half-width of a central 90\% interval of a normal distribution in units of its standard
deviation, to obtain one standard deviation; this assumes Gaussian uncertainties (the effect of
taking them as quoted is given in Sect.~\ref{sec:lfresults}). The Pb+Pb data
are ALICE 0--5\% at 2.76~TeV~\cite{ALICE:2013jfw} and 5.02~TeV~\cite{ALICE:2016fbt}, and CMS 0--5\% at
5.36~TeV~\cite{CMS:2024ykx}. The LHC tables are published symmetrized in $\eta$, so
each $|\eta|$ is used once. The data reach $|\eta|=\etaMaxE$ (ALICE, 2.76~TeV), $\etaMaxF$ (ALICE,
5.02~TeV) and $\etaMaxG$ (CMS, 5.36~TeV). The mean numbers of participants used in
Sect.~\ref{sec:lf}, the collaborations' Glauber-model estimates~\cite{Miller:2007ri}, are $\npartA$, $\npartC$ and $\npartD$ at 19.6, 130 and
200~GeV~\cite{Back:2002wb}, and $\npartE$, $\npartF$ and $\npartG$ at 2.76, 5.02 and
5.36~TeV~\cite{ALICE:2013jfw,ALICE:2016fbt,CMS:2024ykx}.

\paragraph{Identified spectra.} The kernel uses midrapidity $\pi^\pm$, $K^\pm$, $p$ and $\bar p$
spectra for 0--5\% central collisions (the PHOBOS $\dndeta$ are 0--6\%): ALICE at 2.76 and
5.02~TeV~\cite{ALICE:2013mez,ALICE:2019hno} (the 5.02~TeV spectra also serve at 5.36~TeV), PHENIX at
130~\cite{PHENIX:2001vgc} and 200~GeV~\cite{PHENIX:2003iij}, and STAR at
19.6~GeV~\cite{STAR:2017sal}. At 62.4~GeV the STAR spectra~\cite{STAR:2008med} are continued to
higher $p_T$ with STAR~\cite{STAR:2007zea} ($\pi$, $p$) and BRAHMS~\cite{BRAHMS:2009acd} ($K$) data
(0--10\%), scaled to the 0--5\% spectra in the overlap so that only their shapes enter. The STAR spectra alone
at 62.4~GeV~\cite{STAR:2008med} and 200~GeV~\cite{STAR:2003jwm} serve as a variant. The rapidity
dependence of $\langle p_T\rangle$ and of the particle ratios is taken from BRAHMS at
200~GeV~\cite{BRAHMS:2004dwr,BRAHMS:2003wwg}.

\section{Method}
\label{sec:method}

\subsection{The $y\to\eta$ kernel}

For a charged-hadron rapidity density $g(y)=\dndy$,
\begin{equation}
\frac{dN_{\mathrm{ch}}}{d\eta}(\eta)=\sum_{i}\int dp_T\,\phi_i(p_T;y)\,\frac{p}{E}\,x_i(y)\,g(y),
\label{eq:kernel}
\end{equation}
with $y=\sinh^{-1}(p_T\sinh\eta/m_T)$, $m_T=\sqrt{p_T^2+m_i^2}$, $p=p_T\cosh\eta$ and
$E=\sqrt{p^2+m_i^2}$ for a hadron of species $i\in\{\pi,K,p,\mathrm{other}\}$ and mass $m_i$. Here
$\phi_i$ is the normalized $p_T$ spectrum of species $i$ and $x_i$ its share of the charged yield,
normalized to $\sum_i x_i=1$ at every $y$. The measured spectra are used directly inside their $p_T$
range, and a L\'evy--Tsallis fit~\cite{Tsallis:1987eu,STAR:2006nmo},
$dN/dp_T\propto p_T[1+(m_T-m_i)/(nT)]^{-n}$, outside it. Charged hyperons ($\Sigma^\pm$, $\Xi^-$,
$\Omega^-$ and their antiparticles) are given $\fOther\%$ of the charged yield, an estimate from the
measured $\Xi$ and $\Omega$~\cite{ALICE:2013xmt} and $\Lambda$~\cite{ALICE:2013cdo} yields at
2.76~TeV, with $\Sigma^\pm$ estimated from $\Lambda$ through isospin symmetry and an assumed
$\Sigma^0/\Lambda$ ratio; they have the proton spectral
shape and the $\Sigma$ mass, bracketed by variants with $\fOtherLo$ and $\fOtherHi\%$. The rapidity
dependence is modeled as $\phi_i(p_T;y)=\phi_i^{0}(p_T/\lambda_i(y))/\lambda_i(y)$, where $\phi_i^0$
is the midrapidity spectrum. The scale $\lambda_i(y)$ and the ratios $x_i(y)/x_\pi(y)$ are
parametrized as $1+c_iy^2$, fitted to the BRAHMS points at 200~GeV in units of $y/\ybeam$, where
$\ybeam=\cosh^{-1}(\snn/2m_p)$ is the beam rapidity, and held at their values at the largest measured
$|y|$ beyond. For protons $\lambda_p(y)$ is not taken from the BRAHMS proton spectra, which cover
four rapidity slices~\cite{BRAHMS:2003wwg}; the kaon curve is used.

\begin{figure}
\includegraphics[width=\columnwidth]{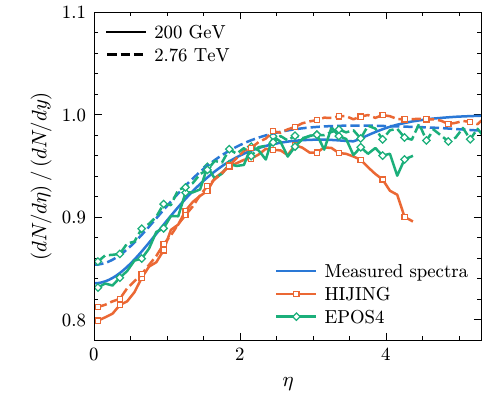}
\caption{Ratio $(dN/d\eta)/(dN/dy)$ for a flat $\dndy$ folded through the kernel built from measured
identified-hadron spectra (blue, lines only) and through the $\eta$-versus-$y$ maps of HIJING
(orange, squares) and EPOS4 (green, diamonds), at 200~GeV (solid) and 2.76~TeV (dashed); the
generator curves are shown for
$|\eta|\le\ybeam-1$. The change of slope of the measured-spectra kernel near $\eta=3.6$ at 200~GeV
marks where the measured BRAHMS rapidity dependence ends and is held constant}
\label{fig:jacobian}
\end{figure}

The kernel variants are: no rapidity dependence; the BRAHMS dependence in absolute $y$; an
$m_T$-exponential low-$p_T$ extrapolation; a $\pm\ptScaleVar\%$ global $p_T$ scale (the uncertainty
of ALICE's pion $\langle p_T\rangle$~\cite{ALICE:2013mez}); the hyperon share; at RHIC a
$\pm\pShareVar\%$ proton share; and the STAR-only spectra at 62.4 and 200~GeV. At 2.76~TeV the kernel's integrated yields ($\pi$ $\yieldPi$, $K$ $\yieldK$, $p$
$\yieldP$) reproduce ALICE's published values ($\yieldPubPi$, $\yieldPubK$,
$\yieldPubP$)~\cite{ALICE:2013mez}. ALICE obtained its own $\dndy$ from its $\dndeta$ with a
Jacobian built from midrapidity identified spectra (at 5.02~TeV those measured at
2.76~TeV)~\cite{ALICE:2013jfw,ALICE:2016fbt}. The ratio of the implied $\dndy$ to it is
\rng{\aliceRatioLoE}{\aliceRatioHiE} at 2.76~TeV and \rng{\aliceRatioLoF}{\aliceRatioHiF} at 5.02~TeV,
a consistency check between two conversions rather than a comparison with a measured $\dndy$.
Figure~\ref{fig:jacobian} compares the effective Jacobian of the kernel with those of the two event
generators used in Sect.~\ref{sec:closure}. At midrapidity EPOS4, which has radial flow, agrees with
the measured-spectra kernel within $\jacEpoLo$ to $+\jacEpoHi\%$ at 200~GeV and 2.76~TeV, whereas
HIJING, with softer spectra, lies \rng{\jacHijAbsLo}{\jacHijAbsHi}$\%$ lower and so produces a deeper kinematic
dip.

\subsection{Hypotheses}

The null hypothesis H0 is a $\dndy$ that is symmetric and non-increasing in $|y|$. Two families
define it: a double Gaussian (dG), $A[G(y-t\sigma)+G(y+t\sigma)]$ with $G$ a Gaussian of width
$\sigma$ and $0\le t\le1$, which keeps it unimodal, and a smeared box (box), $A[L_a(y-Y)-L_a(y+Y)]$
with $L_a(x)=1/(1+e^{x/a})$. A generalized Gaussian (gG), $A\exp[-(|y|/\sigma_g)^\beta]$, serves as a
robustness check. The alternative H1 multiplies the H0 shape by $1-d\exp[-y^2/(2w^2)]$, a
Gaussian dip of depth $0\le d\le1$ and width $w>0$. Apart from small lower limits on widths and on
the exponent $\beta$ that keep the fits numerically defined, and $y_e\ge0$ below, the only parameter
limits are those that define the hypotheses and positivity. Fits are multistart minimizations with
MINUIT~\cite{James:1975dr} through iminuit~\cite{iminuit}; the lowest $\chi^2$
is kept, and H0 is refitted from the H1 solution, so that $\chi^2_{\mathrm{H0}}$ cannot be stuck
above a minimum that H1 has found. Each family also exists with a separate kinematic edge
(dG+edge, box+edge, gG+edge), the shape multiplied by $L_{a_e}(|y|-y_e)$, so that the fall-off near
beam rapidity is not tied to the central shape. These \emph{edge families}, as opposed to the smooth
families above, are used for the limiting-fragmentation test (Sect.~\ref{sec:lf}), the closure tests
(Sect.~\ref{sec:closure}) and a check made after the results (Sect.~\ref{sec:results}); they are
not part of the decision rule.

At RHIC the fit range of the dip test is $|\eta|\le\ybeam$, with $\ybeam=\ybeamA$, $\ybeamB$,
$\ybeamC$ and $\ybeamD$ at 19.6, 62.4, 130 and 200~GeV. Near and beyond beam rapidity the kernel's
rapidity dependence, held constant beyond the BRAHMS range, is least reliable, so PHOBOS points
beyond $\ybeam$ are not used in the dip test. The full range and $|\eta|\le\ybeam-1$ are variants. The LHC data end well before beam rapidity and are used in
full.

\subsection{Uncertainties}

The PHOBOS systematic uncertainties are not split into normalization, uncorrelated and shape
parts. At RHIC the covariance is therefore
\begin{multline}
C_{jk}=\big(\sigma^{\mathrm{stat}}_j\big)^2\delta_{jk}+\sigma^{\mathrm{sys}}_j\sigma^{\mathrm{sys}}_k
\big[f\\
+(1-f)\big(u\,\delta_{jk}+(1-u)\,e^{-|\eta_j-\eta_k|/\ell}\big)\big],
\end{multline}
with $\sigma^{\mathrm{stat}}_j$ and $\sigma^{\mathrm{sys}}_j$ the statistical and systematic
uncertainties of point $j$. Here $f$ is the fully correlated (normalization) share of the systematic
variance; of the remainder, a share $u$ is uncorrelated from point to point and $1-u$ is correlated
with length $\ell$ in $\eta$. $\nCovScan$ combinations are scanned: $f\in\{0,0.5,0.8\}$ with $u=1$, or with
$u\in\{0.5,0.2\}$ and $\ell\in\{0.5,1.5\}$; $f=u=\ell=0.5$ is nominal. At the LHC the nominal
covariance is built from the systematic components documented in each publication
(\ref{app:cov}); its documented variants and the scan serve as variants. At nominal settings
$\chi^2_{\mathrm{H0}}/\mathrm{ndf}$ of the decisive family (Sect.~\ref{sec:decision}) is
\rng{\chiNdfNomRHIClo}{\chiNdfNomRHIChi} at RHIC and
\rng{\chiNdfNomLHClo}{\chiNdfNomLHChi} at the LHC: the published uncertainties are large compared with
the point-to-point scatter of the data. At RHIC the more strongly correlated settings
of the scan reach $\chi^2_{\mathrm{H0}}/\mathrm{ndf}=\chiNdfScanRHIChi$, and the dip-test results below
are checked over the whole scan.

\subsection{Test statistic, calibration and decision rule}
\label{sec:decision}

The test statistic is $q=\chi^2_{\min}(\mathrm{H0})-\chi^2_{\min}(\mathrm{H1})\ge0$, with
$\chi^2_{\mathrm{H0}}\equiv\chi^2_{\min}(\mathrm{H0})$. Its distribution
under H0 is obtained from $\nToys$ pseudo-experiments per configuration, drawn from the best-fit H0
prediction with covariance $C$ and refitted exactly as the data; $p$ is the fraction of
pseudo-experiments with $q\ge q_{\mathrm{obs}}$, and $q_{95}$, the 95th percentile, is
\rng{\qNinetyFiveLo}{\qNinetyFiveHi} at nominal settings. The decision at each energy is drawn from
the \emph{decisive family}, the more flexible of the two H0 families (the one with the lower
$\chi^2_{\mathrm{H0}}$ at nominal settings), because testing against a rigid null family would
mistake its rigidity for evidence of a dip. A dip is \emph{required} if $p<0.0027$ at nominal settings and $p<0.05$ in every
kernel and covariance variant; it is \emph{not required} if $p>0.05$ at nominal settings; anything
else is inconclusive. With $\nToys$ pseudo-experiments the smallest resolvable $p$ is $\pRes$, above
the threshold of 0.0027, so any configuration of the decisive family with $p<0.05$ was to be rerun
with more pseudo-experiments before a decision. The $p$-values are those of the selected family; the
selection is not repeated in the pseudo-experiments. At nominal settings both families give
$p\ge\pNomBothMin$ at every energy, so the decision does not depend on the selection.

\subsection{Sensitivity}
\label{sec:sensitivity}

Dips of depth $d$ and width $w$ are injected into the best-fit H0 shape of the decisive family, on
a grid of $\nWidths$ widths ($w=\widthList$) and $\nDepths$ depths from $\dminGridLo$ to
$\dmaxInj$. Pseudo-experiments are generated from each and analyzed with the same procedure, and
$d_{\min}(w)$ is the depth at which half of them give $q>q_{95}$, i.e.\ $p<0.05$, interpolated
linearly in $d$. $d_{\min}$ therefore refers to a truth that the null family describes exactly,
folded with the nominal kernel; it does not include kernel or shape misspecification. With $\nToys$
pseudo-experiments per point, a rejection rate of 50\% has a binomial uncertainty of $\binomHalf$.
At the smallest injected depth, 1\%, the rejection rate is \rng{\lowRateLo}{\lowRateHi} (0.05 without a
dip, binomial uncertainty 0.015); it is largest for the narrowest widths at the LHC, where $d_{\min}$ is
smallest, so a 1\% dip is already partly detected there, but statistical scatter is not separated from
this.

\section{Results on the midrapidity dip}
\label{sec:results}

\begin{figure*}
\includegraphics[width=\textwidth]{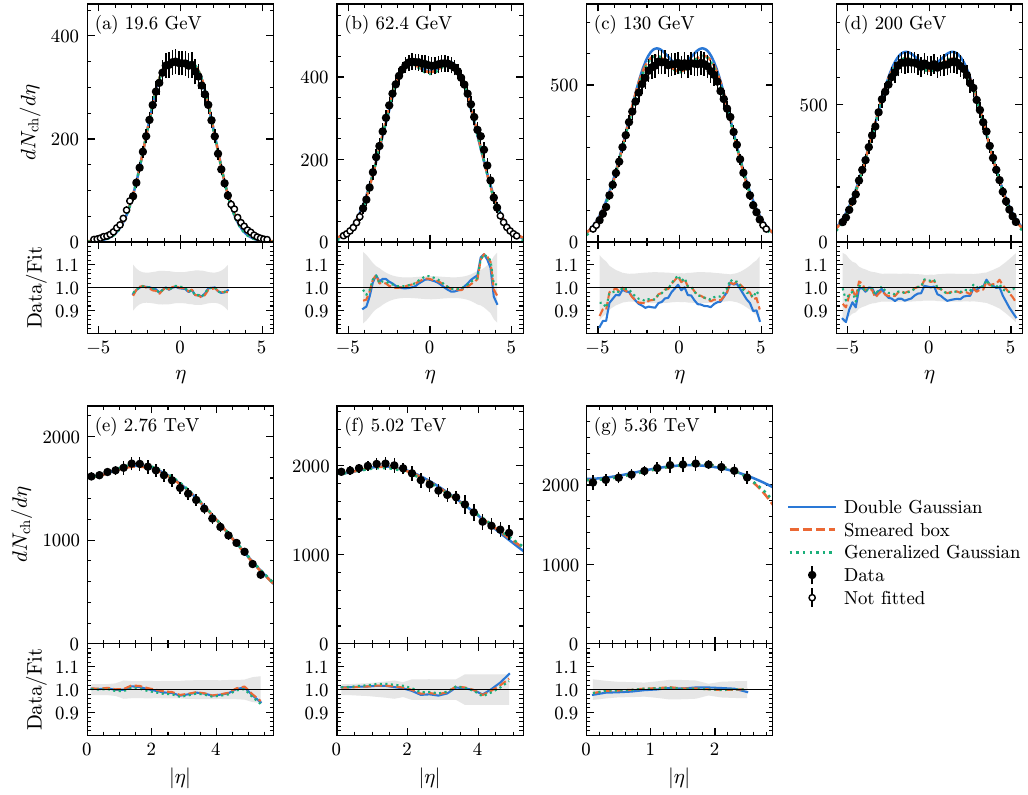}
\caption{Charged-hadron pseudorapidity density in the most central collisions:
(a)--(d) Au+Au (PHOBOS~\cite{Back:2002wb,PHOBOS:2005zhy,PHOBOS:2010eyu}, 0--6\%) at $\snn=19.6$,
62.4, 130 and 200~GeV; (e)--(g) Pb+Pb (ALICE~\cite{ALICE:2013jfw,ALICE:2016fbt},
CMS~\cite{CMS:2024ykx}, 0--5\%) at 2.76, 5.02 and 5.36~TeV. Points: data with total uncertainties.
Lines: best fits of
dip-free $\dndy$ shapes (solid: double Gaussian; dashed: smeared box; dotted: generalized
Gaussian) folded to $\eta$ with the measured-spectra kernel at nominal settings. Lower panels:
Data/Fit; shaded band: relative data uncertainty. RHIC fits use $|\eta|\le\ybeam$; points beyond
(open circles) are not fitted. The LHC
distributions are published symmetrized, so each $|\eta|$ is used once}
\label{fig:fits}
\end{figure*}

\begin{figure*}
\includegraphics[width=\textwidth]{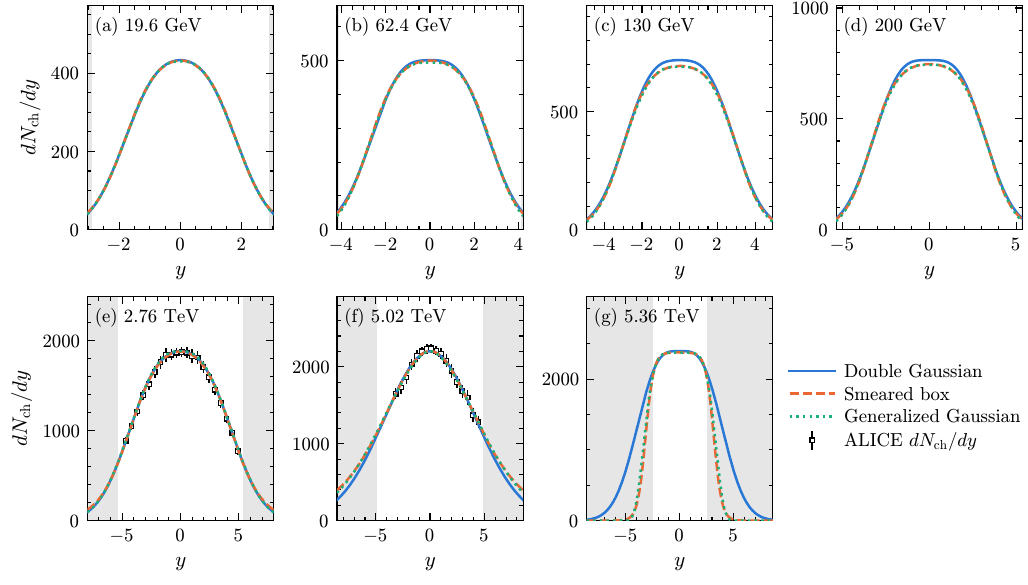}
\caption{Charged-hadron rapidity densities implied by the fits of Fig.~\ref{fig:fits}, panels as in
Fig.~\ref{fig:fits}, with the $\dndy$ obtained by ALICE from its $\dndeta$ at 2.76 and
5.02~TeV~\cite{ALICE:2013jfw,ALICE:2016fbt} (open squares, every second point). The horizontal axes
span $\pm\ybeam$. Shaded: rapidities beyond the largest $|\eta|$ used in the fit, where the curves are
extrapolations of the assumed shape}
\label{fig:dndy}
\end{figure*}

\begin{figure*}
\includegraphics[width=\textwidth]{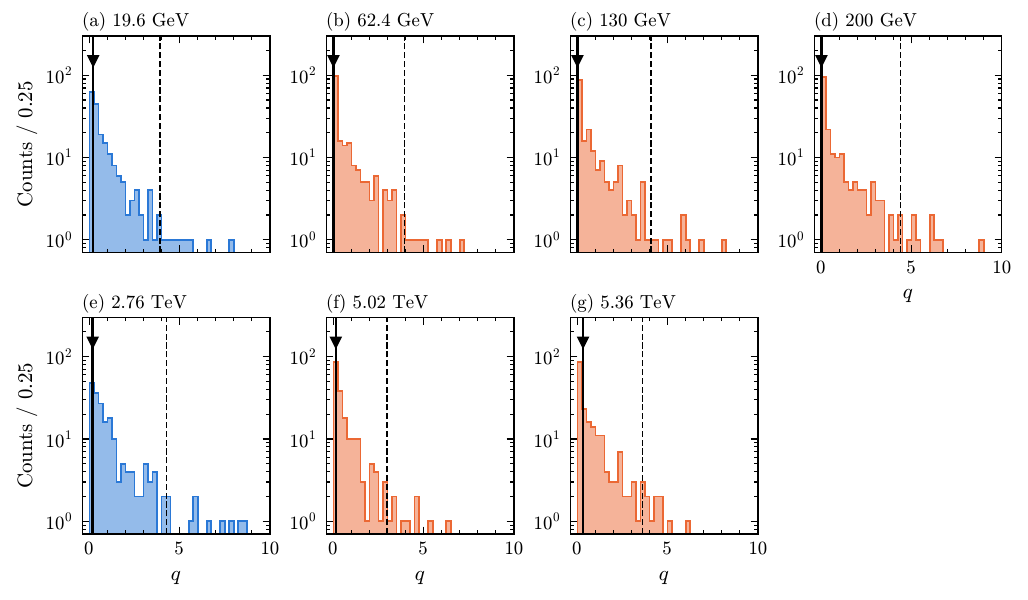}
\caption{Distribution of the test statistic $q$ under the dip-free hypothesis ($\nToys$
pseudo-experiments per panel) for the decisive shape family at nominal settings (blue: double
Gaussian; orange: smeared box; Table~\ref{tab:decision}), panels as in Fig.~\ref{fig:fits}. Solid
line and triangle: observed $q$ (Table~\ref{tab:decision}); dashed line: 95th percentile}
\label{fig:pvalues}
\end{figure*}

Figure~\ref{fig:fits} shows the data with the best dip-free fits and Fig.~\ref{fig:dndy} the implied
rapidity densities. All three families describe the data at every energy. Table~\ref{tab:decision}
summarizes the test, which covers $\nConfAll$ configurations with $\nToys$ pseudo-experiments each.
At every energy the decision is ``dip not required''. The smallest $p$ of the decisive family over
all kernel, range and covariance variants is $\pMinAll$, and no configuration of a decisive family
has $p<0.05$. The distributions of $q$ are shown in Fig.~\ref{fig:pvalues}.

\begin{table*}
\caption{Results of the dip test. Family: decisive H0 family (lower $\chi^2_{\mathrm{H0}}$ of the
double Gaussian and the smeared box at nominal settings). $\chi^2_{\mathrm{H0}}/\mathrm{ndf}$,
$q_{\mathrm{obs}}$, $q_{95}$ and $p$: decisive family, nominal kernel and covariance; $p_{\min}$:
smallest $p$ of the decisive family over all kernel, range and covariance variants. $d_{\min}$ (in
percent): depth at which 50\% of the pseudo-experiments give $p<0.05$ at the given width;
``$>\dmaxInjPct$'': not reached within the injected depths. The decision is ``dip not required'' at
every energy}
\label{tab:decision}
\centering\small
\begin{tabular}{lccccccc}
\hline\noalign{\smallskip}
$\sqrt{s_{NN}}$ & family & $\chi^2_{\mathrm{H0}}/\mathrm{ndf}$ & $q_{\mathrm{obs}}$ & $q_{95}$ & $p$ & $p_{\min}$ & $d_{\min}$ (\%), $w=0.5/1/2$ \\
\noalign{\smallskip}\hline\noalign{\smallskip}
19.6~GeV & dG & 1.4/27 & 0.21 & 3.9 & 0.75 & 0.23 & 15.8 / $>25$ / $>25$ \\
62.4~GeV & box & 11.3/39 & 0.00 & 3.9 & 1.00 & 0.81 & 6.7 / 12.5 / $>25$ \\
130~GeV & box & 14.5/47 & 0.00 & 4.1 & 1.00 & 0.74 & 7.9 / 11.6 / $>25$ \\
200~GeV & box & 7.6/51 & 0.00 & 4.4 & 1.00 & 0.75 & 7.4 / 9.6 / 24.4 \\
2.76~TeV & dG & 6.7/19 & 0.18 & 4.3 & 0.84 & 0.43 & 3.6 / 7.3 / 17.9 \\
5.02~TeV & box & 1.9/17 & 0.13 & 3.0 & 0.66 & 0.23 & 4.0 / 8.0 / $>25$ \\
5.36~TeV & box & 0.4/10 & 0.32 & 3.6 & 0.54 & 0.24 & 6.1 / 8.2 / 21.6 \\
\noalign{\smallskip}\hline
\end{tabular}

\end{table*}

\begin{figure}
\includegraphics[width=\columnwidth]{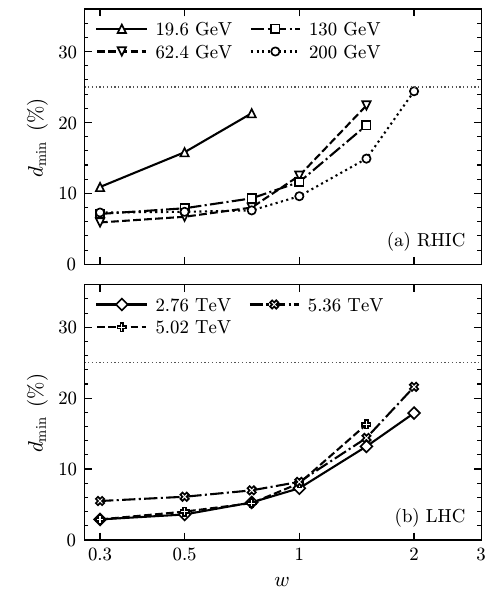}
\caption{Smallest dip depth rejected at the 5\% level with 50\% probability, $d_{\min}$, versus the
dip width $w$ (in units of rapidity) for the decisive shape family: (a) RHIC, (b) LHC energies.
Dotted line: largest injected depth; widths at which $d_{\min}$ exceeds it are not shown}
\label{fig:sensitivity}
\end{figure}

``Not required'' is not ``excluded''. Figure~\ref{fig:sensitivity} shows the sensitivity. For
$w\le1$, $d_{\min}$ is \rng{\dminLHClo}{\dminLHChi}\% at the LHC and \rng{\dminRHIClo}{\dminRHIChi}\% at
62.4--200~GeV: deeper dips would have given $p<0.05$ in at least half of the pseudo-experiments.
At 19.6~GeV, where the fit range is shorter and the uncertainties are larger, $d_{\min}$ is reached
only for $w\le\wMaxA$, with values of \rng{\dminAlo}{\dminAhi}\%. Broad depressions are only weakly
constrained, because the dip-free families can absorb them: at $w=2$ a dip within the injected range
is detectable only at \dminWtwoE, with $d_{\min}$ of \rng{\dminWtwoLo}{\dminWtwoHi}\%.

The only configurations with $p<0.05$ belong to the double-Gaussian family at 62.4--200~GeV, where it
is not the decisive family, under covariance settings that leave little freedom for correlated shape
errors: a normalization share $f=0.8$, or $f=0.5$ with an uncorrelated remainder ($u=1$); there $q$
reaches $\qDgMax$, with $p<\pRes$ in the most extreme settings. Its fits sit at the unimodal
boundary $t=1$, and its best ``dip'' is broader than the distribution itself. We interpret it as a
reshaping of the whole distribution that the family cannot produce, rather than as a midrapidity
structure. This is the
situation the choice of the more flexible family was made for. As a check made after the results,
the edge families (Sect.~\ref{sec:method}) were fitted to the RHIC data. They describe the data
better than the smooth families do: at nominal settings $\chi^2_{\mathrm{H0}}$ of the smeared box falls from
\rng{\chiBoxLo}{\chiBoxHi} to \rng{\chiBoxELo}{\chiBoxEHi} at 62.4--200~GeV, suggesting that the data fall
off faster near beam rapidity than the smooth shapes allow. They find no dip either: $q\le\qEdgeMax$ at
the four RHIC energies at nominal settings, and $q\le\qEdgeMaxAll$ at the covariance setting where the
double Gaussian gives its largest $q$. The test is one-sided: at RHIC the data have slightly less $\eta$-dip than the
best dip-free prediction, within the uncertainties, and an excess at midrapidity is not tested.

Three further checks of the dip test were made after the results and are not part of the decision
rule (\ref{app:amendments}). First, PHOBOS corrections are presumably the same at $+\eta$
and $-\eta$, which the shape term of Sect.~\ref{sec:method} does not assume; with the shape
correlation taken as a function of $||\eta_j|-|\eta_k||$ instead, at nominal $f$, $u$ and $\ell$, the
RHIC data give $p\ge\mirrorPmin$, and $d_{\min}$ for $w\le1$ at 62.4--200~GeV becomes
\rng{\mirrorDminLo}{\mirrorDminHi}\% (with 60 pseudo-experiments per depth), against
\rng{\dminRHIClo}{\dminRHIChi}\% at nominal settings. Second, $d_{\min}$ refers to a truth that the null
family describes. As a truth outside the families, the best fit of the smeared box with a kinematic
edge at 200~GeV and 2.76~TeV, times a dip, was folded with the nominal kernel and fitted, without
fluctuations, with the decisive family. Without a dip it gives $q\le\outsideQzeroMax$, so the shape
difference alone does not produce a false dip; the depth at which $q$ exceeds $q_{95}$, at $w=0.5$
and 1, is \rng{\outsideDLo}{\outsideDHi}\% on the injection grid, \rng{\outsideRatioLo}{\outsideRatioHi}
times $d_{\min}$. A dip in a truth that the families do not describe must therefore be about twice
as deep to be detected. Third, 95\% confidence-level upper limits on $d$ were computed with the
CL$_s$ method~\cite{Read:2002hq}, with $q$ as the test statistic and the stored injection
pseudo-experiments at nominal settings: $\mathrm{CL}_s(d)=P(q\le q_{\mathrm{obs}}\,|\,d)/P(q\le
q_{\mathrm{obs}}\,|\,0)$, with the limit where it falls to 0.05. They are \rng{\clsLoHalf}{\clsHiHalf}\%
at $w=0.5$ and \rng{\clsLoOne}{\clsHiOne}\% at $w=1$ from 62.4~GeV to 5.36~TeV; at 19.6~GeV the
limit is \clsAHalf\% at $w=0.5$ and is not reached within the injected depths at $w=1$. Like
$d_{\min}$, they hold for a truth described by the null family and the nominal kernel.

\section{Closure tests with event generators}
\label{sec:closure}

The kernel relies on two approximations, the factorized rapidity dependence of the spectra and the
extrapolation of the measured $p_T$ ranges. Both, and the dip test itself, are tested in two event
generators in which every final-state momentum is known: HIJING~\cite{Wang:1991hta,Gyulassy:1994ew},
without collective flow, and EPOS4~\cite{Werner:2023zvo,Werner:2023fne,Werner:2023mod,Werner:2023jps},
whose core--corona procedure gives a viscous-hydrodynamic core with radial flow, followed by the UrQMD
hadronic cascade~\cite{Bass:1998ca,Bleicher:1999xi}. Both were run at 19.6, 62.4 and 200~GeV and at
2.76 and 5.02~TeV, for impact parameters
$b\le3.3$~fm (Au+Au) and $3.5$~fm (Pb+Pb), with $\pi^0$ and weak decays switched off and spectator
nucleons removed. HIJING (through AliGenHijing of AliPhysics vAN-20260817\_O2-1; nuclear shadowing on,
jet quenching off;
spectators removed with AliGenHijing's rule) has $\hijNevHigh$ events per energy
at 200~GeV and above and $\hijNevLow$ below; its final-state energy equals the participant energy
$\Npart\snn/2$ with a relative deviation below $\hijEnergyDev$. EPOS4 (version 4.0.3, with the
lead--lead example options of the EPOS4 authors: core--corona procedure, hydrodynamics with the
x3ff equation of state, and the hadronic cascade) has at least $\epoNevHigh$ events per energy at
200~GeV and above and $\epoNevLow$
below; the never-interacting spectator nucleons are identified in its event record, and the
remaining final-state energy equals the participant energy within $\epoEnergyDev\%$ per event
($\epoRejected$ event with an energy violation was rejected). In these EPOS4 samples baryon number is
not conserved exactly event by event; this does not enter the tests, because the energy rule takes
its net baryons from data.

\begin{table*}
\caption{Generator closure. Kernel bias: $\sum_j[(\mathrm{fold}_j/\mathrm{true}_j-1)/\epsilon_j]^2$ over
the points of the real data, with $\epsilon_j$ the relative uncertainty of the data, and the
largest $|\mathrm{fold}_j/\mathrm{true}_j-1|$, for the kernel rebuilt from the generator as from data
(Sect.~\ref{sec:closure}; at the LHC with the generator's 200~GeV rapidity dependence carried in
$y/\ybeam$). Dip test: generator pseudo-data with the real data's uncertainties (noise-free,
so $q_{\mathrm{A}}$ approximates the median $q$) for the decisive family of these pseudo-data,
which can differ from Table~\ref{tab:decision}, against $q_{95}$ of the real-data
pseudo-experiments for that family. $R$: energy estimate of Sect.~\ref{sec:lf} (nominal rule) divided
by the participant energy, within the measured $\eta$ range up to $\ybeam$ (as for the calibration)
and over $|y|\le\ybeam$ (as for the extrapolation at the LHC)}
\label{tab:closure}
\centering\small
\begin{tabular}{lllcccccc}
\hline\noalign{\smallskip}
 & & & \multicolumn{2}{c}{kernel bias} & \multicolumn{2}{c}{dip test} & \multicolumn{2}{c}{$R$ (energy estimate)} \\
generator & $\sqrt{s_{NN}}$ & family & $\chi^2$/points & max.\ dev.\ (\%) & $q_{\mathrm{A}}$ & $q_{95}$ & measured range & $|y|\le y_{\mathrm{beam}}$ \\
\noalign{\smallskip}\hline\noalign{\smallskip}
HIJING & 19.6~GeV & dG & 14.9/30 & 9.8 & 0.01 & 3.90 & 1.09 & 1.12 \\
HIJING & 62.4~GeV & dG & 9.7/42 & 10.4 & 0.10 & 4.09 & 1.03 & 1.04 \\
HIJING & 200~GeV & box & 11.5/54 & 10.0 & 12.71 & 4.39 & 0.99 & 1.00 \\
HIJING & 2.76~TeV & box & 0.5/22 & 1.8 & 0.00 & 4.46 & -- & 1.04 \\
HIJING & 5.02~TeV & box & 0.1/20 & 0.7 & 0.00 & 2.98 & -- & 1.05 \\
\noalign{\smallskip}\hline\noalign{\smallskip}
EPOS4 & 19.6~GeV & box & 17.7/30 & 19.1 & 0.22 & 3.94 & 1.22 & 1.24 \\
EPOS4 & 62.4~GeV & box & 15.6/42 & 36.2 & 0.00 & 3.95 & 1.15 & 1.17 \\
EPOS4 & 200~GeV & dG & 9.1/54 & 29.1 & 5.81 & 4.11 & 1.11 & 1.12 \\
EPOS4 & 2.76~TeV & dG & 0.4/22 & 1.1 & 0.01 & 4.27 & -- & 1.28 \\
EPOS4 & 5.02~TeV & box & 0.3/20 & 0.9 & 0.24 & 2.98 & -- & 1.31 \\
\noalign{\smallskip}\hline
\end{tabular}

\end{table*}

Folding each generator's $\dndy$ with its own $\eta$-versus-$y$ map reproduces its $\dndeta$ within
$\hijExactMax\%$ (HIJING) and $\epoExactMax\%$ (EPOS4). The kernel is then rebuilt from the
generator as from data: its midrapidity spectra, restricted to the $p_T$ ranges measured at
each energy and extrapolated in the same way, and its rapidity dependence fitted within the BRAHMS
range and held constant beyond. At RHIC this is the generator's own rapidity dependence at each
energy; at the LHC the generator's 200~GeV dependence is carried in $y/\ybeam$, as the data kernel
carries BRAHMS's. Folded with the generator's true $\dndy$, this kernel
reproduces its $\dndeta$ at the $\eta$ points of the data within $\hijKernelLHCsc\%$ (HIJING) and
$\epoKernelLHCsc\%$ (EPOS4) at the LHC (Table~\ref{tab:closure}). At RHIC the deviations are largest
at the most forward points, up to $\hijDevRHICmax\%$ in HIJING and $\epoDevRHICmax\%$ in EPOS4; within
$|\eta|\le\ybeam-1$ they are \rng{\hijDevInnerLo}{\hijDevInnerHi}$\%$ and \rng{\epoDevInnerLo}{\epoDevInnerHi}$\%$.
In units of the data uncertainties the mean $\chi^2$ per point is at most $\hijBiasRHICmax$ and
$\epoBiasRHICmax$. With full spectra and each generator's exact rapidity dependence, which leaves only
the factorized form $\phi_i^0(p_T/\lambda_i)/\lambda_i$, the deviations within $|\eta|\le\ybeam-1$ are
\rng{\hijExactInnerLo}{\hijExactInnerHi}$\%$ in HIJING and \rng{\epoExactInnerLo}{\epoExactInnerHi}$\%$ in
EPOS4, and up to $\hijExactAllMax\%$ and $\epoExactAllMax\%$ at all points. In HIJING the factorized $p_T$
scaling itself is therefore violated at this level; in EPOS4 part of the forward deviations comes from
holding the rapidity dependence constant beyond the BRAHMS range.

Restricting the spectra to the narrow STAR $p_T$ ranges raises the extrapolated proton
$\langle p_T\rangle$ at 200~GeV by $\hijPtStarD\%$ in HIJING and by $\epoPtStarD\%$ in EPOS4 (kaons
$\epoPtStarKD\%$), whereas the nominal ranges change it by $\hijPtNomD\%$ and $\epoPtNomD\%$. The
generator with radial flow thus shows a bias of the same sign as the data, in which the STAR-only
spectra give a proton $\langle p_T\rangle$ higher by $\mptBiasData\%$ than the nominal spectra at
200~GeV (Sect.~\ref{sec:discussion}); HIJING, with softer spectra, shows it only weakly.

The dip test applied to noise-free generator pseudo-data with the uncertainties of the real data,
whose $q$ (denoted $q_{\mathrm{A}}$) approximates the median $q$, finds no dip
at four of the five energies in both generators. At 200~GeV both give a false dip, $q_{\mathrm{A}}$
above $q_{95}$: $q_{\mathrm{A}}=\hijFalseQ$ against $\hijFalseQnf$ in HIJING (decisive family
\hijFalseFam) and $q_{\mathrm{A}}=\epoFalseQ$ against $\epoFalseQnf$ in EPOS4 (\epoFalseFam;
Table~\ref{tab:closure}). In EPOS4 it is driven by the most forward
points: restricting the fit to
$|\eta|\le\ybeam-1$ removes it ($q_{\mathrm{A}}=\epoYbmQ$), as do the edge families ($\epoEdgeQ$). In HIJING,
whose $\dndy$ has a distinct forward component, neither does ($q_{\mathrm{A}}=\hijYbmQ$ and $\hijEdgeQ$); only
restricting the fit to $|\eta|\le3$ removes it ($q_{\mathrm{A}}=\hijEtaThreeQ$; EPOS4 $\epoEtaThreeQ$).
With each generator's own $\eta$-versus-$y$ map instead of the rebuilt kernel, $q_{\mathrm{A}}=\hijExactQ$
in HIJING, still above $q_{95}$, and $\epoExactQ$ in EPOS4, just below it: the false dip in HIJING comes
from the dip-free families' failure to describe its forward $\dndy$, and in EPOS4, where $q_{\mathrm{A}}$
exceeds $q_{95}$ only marginally, partly from the kernel. HIJING's true $\dndy$ itself has a midrapidity
dip of $\hijTrueDip\%$, far below the sensitivity at this energy. The
procedure can therefore produce a false dip at top RHIC energy, and a positive result would need
checks with reduced fit ranges. In the data no dip is found with either the smooth or the edge
families (Sect.~\ref{sec:results}).

\section{Identified and inclusive BRAHMS yields at 200 GeV}
\label{sec:brahms}

\begin{figure}[t]
\includegraphics[width=\columnwidth]{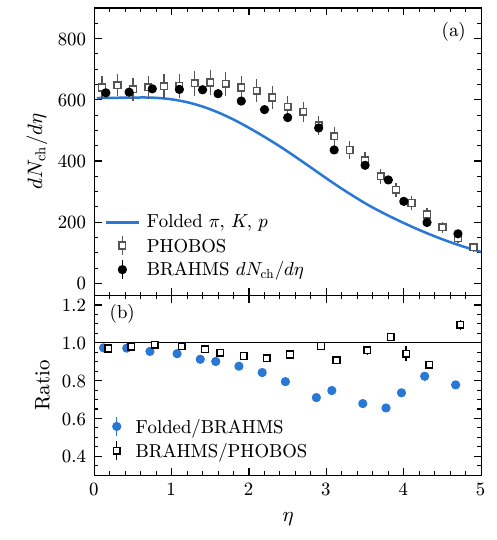}
\caption{Au+Au at $\snn=200$~GeV. (a) Charged-hadron $\dndeta$ measured by
BRAHMS~\cite{BRAHMS:2001llo} (0--5\%, filled circles, statistical uncertainties) and
PHOBOS~\cite{Back:2002wb,PHOBOS:2010eyu} (0--6\%, open squares, total uncertainties,
$\eta\ge0$), and the BRAHMS identified $\pi$, $K$, $p$ rapidity densities~\cite{BRAHMS:2004dwr,BRAHMS:2003wwg}
(Gaussian fits for $\pi$ and $K$; measured points for $p$, held constant beyond the last point)
folded to $\eta$ with the measured-spectra kernel (line; the band of kernel variants is barely wider than the line).
(b) Folded identified yield over the BRAHMS charged $\dndeta$ (filled) and BRAHMS over PHOBOS (open)}
\label{fig:brahms}
\end{figure}

BRAHMS measured both identified $\pi$, $K$ and $p$ rapidity densities and the charged $\dndeta$ at
200~GeV~\cite{BRAHMS:2004dwr,BRAHMS:2003wwg,BRAHMS:2001llo}. Folding the identified rapidity densities
through the kernel (Fig.~\ref{fig:brahms}) gives $\brahmsFoldMid$ of BRAHMS's own $\dndeta$ at
$\eta=\brahmsEtaMid$, falling to $\brahmsFoldMin$ at forward pseudorapidity, for every kernel variant. The
ratio of the BRAHMS and PHOBOS charged distributions is \rng{\brahmsPhobosLo}{\brahmsPhobosHi}. The
identified yields are thus narrower in rapidity than the inclusive charged distribution of the same
experiment, and BRAHMS identified $\dndy$ does not provide a reliable check of an $\eta\to y$
inversion beyond $|y|\approx1$. The origin is not resolved here; it may lie in the $p_T$ extrapolation of the forward
identified spectra, in charged species other than $\pi$, $K$ and $p$, or in the forward systematics
of the inclusive measurement. The kernel uses only the relative rapidity dependence of BRAHMS's
$\langle p_T\rangle$ and particle ratios, not the identified yields; its rapidity dependence is
tested in two generators (Sect.~\ref{sec:closure}) and varied between none, scaled and absolute.

\section{Limiting fragmentation at the LHC}
\label{sec:lf}

Limiting fragmentation is tested in the variable $\eta'=\eta-\ybeam$, in which it predicts
$(2/\Npart)\,\dndeta$ to be independent of energy. The most forward LHC points lie at
$\eta'=\etapMinE$ (2.76~TeV), $\etapMinF$ (5.02~TeV) and $\etapMinG$ (5.36~TeV), so any statement at
the LHC rests on an extrapolation. Only at 2.76~TeV do the data come close to the window
$\lfLo\le\eta'\le0$ of the test below: at the most forward point, $\eta'=\ovEtap$, just outside it,
the ratio to PHOBOS at 200~GeV, interpolated linearly in $\eta'$, is $\ovRatio\pm\ovErr$ (total
uncertainties of both measurements in quadrature, without the $\Npart$ uncertainties).

\subsection{The data alone do not fix the extrapolation}

Dip-free $\dndy$ shapes that describe the 2.76~TeV data equally well give
$(2/\Npart)\,\dndeta={}$\rng{\freeZeroLoE}{\freeZeroHiE} at $\eta'=0$ for the three smooth families and
\freeZeroEdgeE{} for the same families with a kinematic edge (medians of bootstrap fits, i.e.\ refits
of pseudo-data drawn from the best fit with the data covariance, \nBoot{} per family). With the
energy estimator $f_E$ of Sect.~\ref{sec:sumrule}, the best fits of the smooth families give
\rng{\freeEnergyLoE}{\freeEnergyHiE} (2.76~TeV) and \rng{\freeEnergyLoF}{\freeEnergyHiF} (5.02~TeV) times
its RHIC calibration $\kappa$, and those of the edge families \freeEnergyEdgeE{} times $\kappa$ at
2.76~TeV (Fig.~\ref{fig:sumrule}b, where $\kappa$ is the shaded band). A limiting-fragmentation conclusion
drawn from such extrapolations reflects the assumed shape.

\subsection{Energy sum rule}
\label{sec:sumrule}

The estimated energy fraction
\begin{multline}
f_E=\frac{1}{\Npart\snn/2}\int dy\,\cosh y\,\Big[\Big(g(y)-\frac{B(y)}{c_B}\Big)\\
\times e_{\mathrm{prod}}(y)+B(y)\,\langle m_T\rangle_B(y)\Big]
\end{multline}
counts the produced charged hadrons with their neutral partners,
$e_{\mathrm{prod}}(y)=\sum_i x_i(y)\langle m_T\rangle_i(y)\,\nu_i$, where $x_i$ and the mean
transverse masses $\langle m_T\rangle_i$ follow from the kernel and $\nu_i$, the number of charged plus
neutral hadrons per charged hadron, is $\nu_\pi=\Rpi$ and $\nu_K=\nu_p=\nu_{\mathrm{other}}=\Rk$. The
explicit net-baryon component $B(y)$ is normalized to $\Npart$. Its profile and its mean transverse mass $\langle m_T\rangle_B$
come from the BRAHMS net-proton measurement at 200~GeV~\cite{BRAHMS:2003wwg}, with $c_B=\cB$ net
baryons per net proton, and are carried to other energies at fixed rapidity loss, which BRAHMS
found to be the same within uncertainties at 62.4 and 200~GeV~\cite{BRAHMS:2003wwg,BRAHMS:2009wlg}.
This net-baryon profile gives a
mean rapidity loss of $\netbDy$ (BRAHMS: $2.04\pm0.10$) and an energy per net baryon of
\netbE~GeV (BRAHMS: $30\pm2$ and $26\pm5$~GeV from two parametrizations)~\cite{BRAHMS:2003wwg}.
Besides this nominal rule, \nVariantsMinusOneWord{} variants are used: a flat rule ($\nu=\nuFlat$ for
all species and no net-baryon component), a species rule (the factors $\nu_i$, no net-baryon
component), the net-baryon profile shifted away from beam rapidity by $\netbShiftLo$ and
$\netbShiftHi$ units, toward the ends of the conservative range 1.45--2.45 of the rapidity loss
quoted by BRAHMS~\cite{BRAHMS:2003wwg} (mean loss \netbDyLo{} and \netbDyHi{} at 200~GeV after the cut
at beam rapidity), and $\langle m_T\rangle_B$ falling to the proton mass at beam rapidity.

\begin{figure*}
\includegraphics[width=\textwidth]{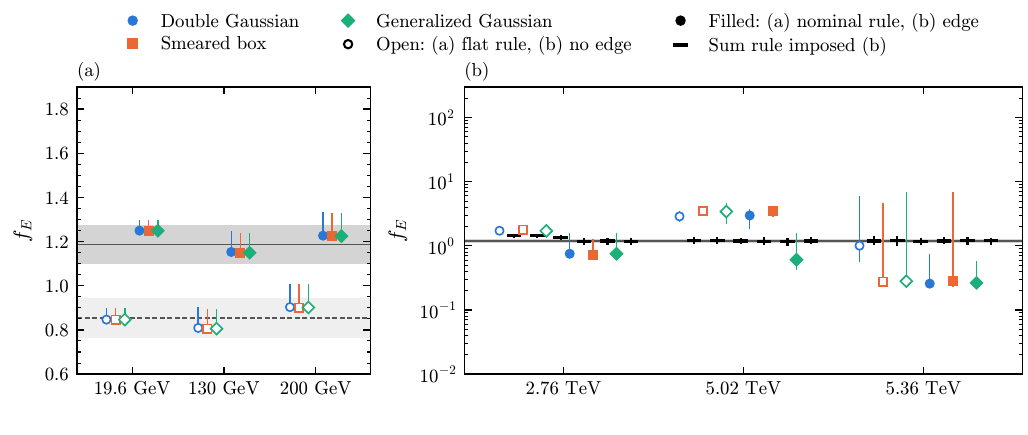}
\caption{Energy sum rule. (a) RHIC: $f_E$ of the dip-free fits with a kinematic edge within
$|\eta|\le\ybeam$, integrated up to the most forward fitted $|\eta|$ (markers: nominal kernel; bars: range over the kernel's
rapidity-dependence variants) for the flat rule (open; $\nu=\nuFlat$ for all species) and the
nominal rule (filled); shaded bands and lines: calibration $\kappa\pm\sigma_\kappa$ from 130 and
200~GeV for each rule (dashed: flat; solid: nominal); 19.6 and 62.4~GeV are checks outside the calibration
(Sect.~\ref{sec:sumrule}). (b) LHC, nominal rule: $f_E$ of
the dip-free fits to the data alone (median and central 95\% of bootstrap fits; open: no edge,
filled: with edge) and with the sum rule imposed (black bars; on this scale they coincide with the
band, and their pulls are given in Table~\ref{tab:lf}); shaded: $\kappa\pm\sigma_\kappa$. Without the
rule the data leave $f_E$ essentially undetermined}
\label{fig:sumrule}
\end{figure*}

$f_E$ is an estimator, not the energy fraction itself: it depends on the composition and
$\langle m_T\rangle$ near beam rapidity, where neither is measured. It is therefore calibrated where
the data cover nearly the whole rapidity range and the constant-rapidity-loss profile applies, at 130
and 200~GeV; 62.4 and 19.6~GeV serve as checks. There the three dip-free
families with a kinematic edge, the shapes used at the LHC, are fitted within $|\eta|\le\ybeam$ as in
the dip test, and $f_E$ is integrated up to the most forward fitted point; at the LHC it is
integrated up to beam rapidity. $\kappa$ is their mean with the nominal kernel. $\sigma_\kappa$
combines the larger of their standard deviation and half the largest deviation from $\kappa$ among
the kernel's rapidity-dependence variants, at least half the difference to the smooth families
(which give $\kappaSmooth$), with the PHOBOS scale uncertainty: $\sigScale\%$ of $\kappa-c$, from the
normalization share of the systematic uncertainty and the $\Npart$ uncertainty. The same scale also
enters the reference window of Sect.~\ref{sec:lftest}; this correlation is neglected, which widens the
intervals in $s$. This gives $\kappa=\kappaNom\pm\kappaErr$ for the nominal rule and
$\kappaFlat\pm\kappaFlatErr$ for the flat rule (Fig.~\ref{fig:sumrule}a). Without the net-baryon
component (species rule) the calibration gives $\kappaSpecies\pm\kappaSpeciesErr$: the produced
hadrons with their neutral partners account for about the participant energy, and the net-baryon
component adds $c=\netbTerm$. Of the species-rule value, \rng{\fwdShareLo}{\fwdShareHi}$\%$ comes from beyond
the last BRAHMS point, where the composition and $\langle m_T\rangle$ are held at their last measured
values; there the kernel's proton yield ($\kernPfwd$ per unit $y$ at $y=4$, 200~GeV) is below the
net-proton density of the model ($\netpFwd$). Why $\kappa$ exceeds unity is therefore not identified:
the calibration absorbs it, and only its transfer to the LHC is taken from the generators. As checks,
at 62.4 and 19.6~GeV, fitted and integrated in the same way, the nominal rule gives $f_E=\fEB$ and
$\fEA$ ($\fEBPull$ and $\fEAPull\,\sigma_\kappa$ from $\kappa$) and the flat rule $\fEBflat$ and $\fEAflat$
($\fEBflatPull$ and $\fEAflatPull\,\sigma_\kappa$), although near SPS energies the rapidity loss still grows
with beam rapidity~\cite{BRAHMS:2003wwg}. The generators,
evaluated in the same way, give $R=\hijRcalA$ at 19.6~GeV against $\hijRcal$ at 200~GeV in HIJING
and $\epoRcalA$ against $\epoRcal$ in EPOS4 (Table~\ref{tab:closure}): their estimator bias changes
more with energy than that of the data. Only the transfer of the calibration from RHIC to the LHC is
therefore taken from the generators, and from both of them, below.

At the LHC the fits are repeated with the term $[(f_E-\kappa)/\sigma_\kappa]^2$ (the squared ``pull'')
added to $\chi^2$. Smooth shapes then describe the data poorly, and a separate fall-off below
beam rapidity is needed, which the edge families of Sect.~\ref{sec:method} provide: they describe
both the data and the sum rule (Fig.~\ref{fig:sumrule}b). In the best fits of the test below
(Table~\ref{tab:lf}) the edge lies close to beam rapidity ($y_e={}$\rng{\yeNearLo}{\yeNearHi},
$a_e={}$\rng{\aeNearLo}{\aeNearHi}), except for \edgeBroadList. There the central
shape itself falls off sharply near beam rapidity, and the edge factor ($y_e={}$\rng{\yeBroadLo}{\yeBroadHi},
$a_e={}$\rng{\aeBroadLo}{\aeBroadHi}) acts as a broad suppression of the forward region.

\subsection{Test of limiting fragmentation}
\label{sec:lftest}

\begin{figure*}
\includegraphics[width=\textwidth]{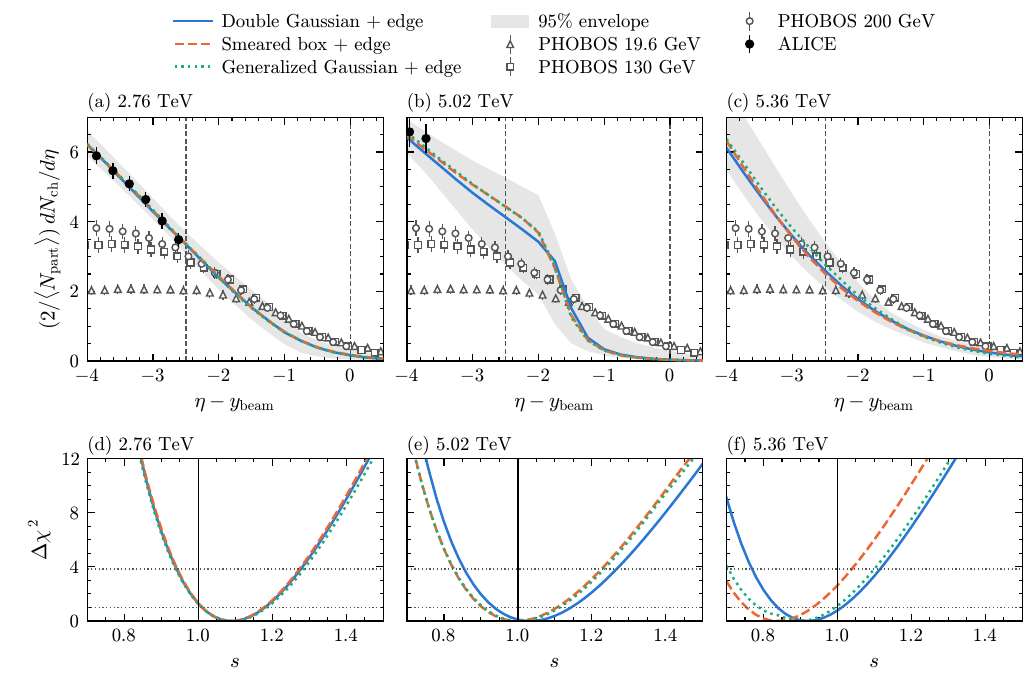}
\caption{Limiting fragmentation. (a)--(c): $(2/\Npart)\,\dndeta$ versus $\eta'=\eta-\ybeam$ for
PHOBOS Au+Au at 19.6, 130 and 200~GeV (open symbols; 62.4~GeV omitted for clarity) and the
ALICE data~\cite{ALICE:2013jfw,ALICE:2016fbt} (filled circles; only the most forward 5.02~TeV points
lie in the plotted range, and none of the CMS~\cite{CMS:2024ykx} 5.36~TeV points). Lines: median LHC
extrapolations allowed by the data and the energy sum rule alone (nominal rule, $\kappa$ as
calibrated) for three dip-free shapes with a kinematic edge; shaded: 95\% envelope of their bootstrap
fits, with $\kappa$ drawn from its uncertainty. Vertical dashed lines: the window
$\lfLo\le\eta'\le0$ of the test. (d)--(f): $\Delta\chi^2$ (data, energy sum rule and imposed
shape) when the PHOBOS 200~GeV distribution scaled by $s$ is imposed in that window; dotted lines
$\Delta\chi^2=1$ and 3.84, vertical line $s=1$}
\label{fig:lf}
\end{figure*}

The sum rule constrains the integral, not the shape, of the fragmentation region: the extrapolations
it selects fall below the PHOBOS data near $\eta'=0$ (Fig.~\ref{fig:lf}). The RHIC
shape is therefore imposed, and only its normalization is tested. The PHOBOS 200~GeV distribution
$(2/\Npart)\,\dndeta$ with $\Npart=\npartPhobos\pm\npartPhobosErr$, scaled by $s$,
is added as pseudo-data at its own points in $\lfLo\le\eta'\le0$. Its covariance is the nominal
RHIC covariance of these points plus the $\Npart$ uncertainty as a common normalization; for $s$
times this distribution the uncertainties scale with $s$ and the covariance with $s^2$. This is
equivalent to comparing the LHC prediction divided by $s$ with the PHOBOS points and their fixed
covariance, so $\chi^2_{\mathrm{pseudo}}$ needs no determinant term. Only the nominal covariances of
the LHC data and of the PHOBOS points are used here; the covariance variants of the dip test are not
propagated. The total $\chi^2_{\mathrm{total}}=\chi^2_{\mathrm{data}}+\mathrm{pull}^2+
\chi^2_{\mathrm{pseudo}}$ is minimized at each $s$ on a grid of step $\fineStep$ for the three edge
families, and limiting fragmentation corresponds to $s=1$. Each scan point is refitted from its
neighbors' solutions until no point improves. Where two solution branches exist, the scan follows
the lower one; the switch between them can produce small kinks in the scan curves. The intervals
use the asymptotic thresholds $\Delta\chi^2=1$ and 3.84; their coverage was not checked with
pseudo-experiments.

\begin{table*}
\caption{LHC/RHIC ratio $s$ in $\lfLo\le\eta'\le0$ ($\eta'=\eta-\ybeam$) with the RHIC shape imposed,
for the three edge families. $\kappa$ as calibrated: best $s$ from a parabola through the scan
minimum (step $\fineStep$), 95\% interval ($\Delta\chi^2<3.84$, asymptotic threshold, ends
interpolated), $\Delta\chi^2$ at $s=1$, and the sum-rule pull $(f_E-\kappa)/\sigma_\kappa$ at the
minimum. $\kappa\times\kTransfer$ and $\kappa\times\kTransferE$: the transfer factors found in
HIJING and EPOS4 ($\sigma_\kappa$ unchanged). $\kappa_{\mathrm{STAR}}$: calibration with the
STAR-only spectra at 200~GeV ($\sigma_\kappa$ unchanged). Scan grids: $s={}$\rng{\fineLo}{\fineHi} for
$\kappa$ as calibrated, \rng{\kapLo}{\kapHi} in steps of $\kapStep$ for the other columns; ``$<$'',
``$>$'': not reached on the grid}
\label{tab:lf}
\centering\small\setlength{\tabcolsep}{4pt}
\begin{tabular}{llcccccccc}
\hline\noalign{\smallskip}
 & & \multicolumn{3}{c}{$\kappa$ as calibrated} & \multicolumn{2}{c}{$\kappa\times1.06$ (HIJING)} & \multicolumn{2}{c}{$\kappa\times1.17$ (EPOS4)} & $\kappa_{\mathrm{STAR}}$ \\
$\sqrt{s_{NN}}$ & family & best $s$ (95\% interval) & $\Delta\chi^2(s{=}1)$ & pull & best $s$ & $\Delta\chi^2(s{=}1)$ & best $s$ & $\Delta\chi^2(s{=}1)$ & best $s$ \\
\noalign{\smallskip}\hline\noalign{\smallskip}
2.76~TeV & dG+edge & 1.09 (0.94--1.28) & 1.3 & 2.5 & 1.10 & 1.5 & 1.11 & 2.0 & 1.10 \\
2.76~TeV & box+edge & 1.09 (0.94--1.27) & 1.3 & 2.5 & 1.09 & 1.5 & 1.10 & 1.9 & 1.10 \\
2.76~TeV & gG+edge & 1.09 (0.94--1.29) & 1.3 & 2.5 & 1.10 & 1.6 & 1.12 & 2.3 & 1.11 \\
5.02~TeV & dG+edge & 1.03 (0.85--1.27) & 0.1 & 2.9 & 1.06 & 0.4 & 1.11 & 1.3 & 1.07 \\
5.02~TeV & box+edge & 1.00 (0.83--1.23) & 0.0 & 2.6 & 1.03 & 0.0 & 1.08 & 0.7 & 1.04 \\
5.02~TeV & gG+edge & 1.00 (0.83--1.23) & 0.0 & 2.6 & 1.03 & 0.1 & 1.09 & 0.8 & 1.04 \\
5.36~TeV & dG+edge & 0.92 (0.77--1.12) & 0.8 & 1.7 & 0.95 & 0.3 & 1.02 & 0.0 & 0.96 \\
5.36~TeV & box+edge & 0.84 (0.68--1.04) & 2.6 & 1.2 & 0.88 & 1.4 & 0.96 & 0.1 & 0.90 \\
5.36~TeV & gG+edge & 0.90 (0.70--1.10) & 1.1 & 1.6 & 0.94 & 0.4 & 1.00 & 0.0 & 0.95 \\
\noalign{\smallskip}\hline
\end{tabular}

\end{table*}

\subsection{Results and the transfer of the calibration}
\label{sec:lfresults}

Table~\ref{tab:lf} gives the results. At 2.76~TeV the ratio is $s={}$\rng{\sLoE}{\sHiE} with
$\Delta\chi^2(s{=}1)={}$\rng{\dOneLoE}{\dOneHiE}; the direct ratio at the most forward data point,
$\ovRatio\pm\ovErr$, is consistent with it. At 5.02~TeV $s={}$\rng{\sLoF}{\sHiF}, with its minimum at
$s\approx1$. At 5.36~TeV, where CMS covers only $|\eta|\le\etaMaxG$ and the result is an
extrapolation by the sum rule, $s={}$\rng{\sLoG}{\sHiG} with $\Delta\chi^2(s{=}1)={}$\rng{\dOneLoG}{\dOneHiG}. The
95\% intervals, taken over the families, are wide, for example \rng{\sNFLoE}{\sNFHiE} at 2.76~TeV and
\rng{\sNFLoF}{\sNFHiF} at 5.02~TeV, because the RHIC uncertainties in the window are large and partly
correlated. The
difference between 5.02 and 5.36~TeV, close in energy, is within these intervals and indicates the
size of the extrapolation uncertainty at 5.36~TeV, where the data end furthest from the window. Over
the \nVariantsWord{} sum rules (coarse scan, step $\coarseStep$) the best $s$ changes by
\varSpreadText. The
coarse grid underestimates $\Delta\chi^2(s{=}1)$ slightly; on it the largest value is $\dOneVarHiE$
at 2.76~TeV and $\dOneVarHiG$ at 5.36~TeV (\varMaxRuleG{} rule, \varMaxFamG), \varEdgeText. The sum-rule pull
at the best $s$ is \rng{\pullLoE}{\pullHiE} at 2.76~TeV, \rng{\pullLoF}{\pullHiF} at 5.02~TeV and
\rng{\pullLoG}{\pullHiG} at 5.36~TeV: the LHC fits prefer more estimated energy than the RHIC
calibration allows.

The covariances matter more. The scan was repeated on a grid of step 0.05, each point restarted from
the nominal solutions, for the documented variants of the LHC covariances (and, at 5.02~TeV, the
forward ALICE uncertainties correlated as at 2.76~TeV), for three settings of the covariance scan of
Sect.~\ref{sec:method} for the PHOBOS points in the window, $(f,u,\ell)=(0,1,-)$, $(0.8,0.2,1.5)$ and
$(0.5,0.2,1.5)$, and for the kernel's rapidity dependence at the LHC (none, in absolute $y$, and the
BRAHMS dependence at a fixed distance from beam rapidity). Over these variants the best $s$ is
\rng{\sSystLoE}{\sSystHiE} at 2.76~TeV, \rng{\sSystLoF}{\sSystHiF} at 5.02~TeV and \rng{\sSystLoG}{\sSystHiG} at
5.36~TeV, and $s=1$ leaves the 95\% interval \systOutText. No single setting excludes $s=1$ at all
three energies, and the interval over the families excludes it only \systUnionText. The unpublished correlations of the PHOBOS
uncertainties in the window move the best $s$ by up to $\systShiftMax$, comparable to the half-width of
the 95\% interval. In
\covToys{} pseudo-experiments generated at $s=1$ for box+edge at 2.76 and 5.36~TeV, $\Delta\chi^2(s{=}1)$
exceeds 3.84 in a fraction \rng{\covFracLo}{\covFracHi} of them, consistent with the nominal 0.05
(at 5.36~TeV the minimum lies at the edge of the pseudo-experiments' scan in up to \covEdgeMax{} of
them, for which $\Delta\chi^2(s{=}1)$ is a lower bound). Taking the PHOBOS uncertainties as quoted, at 90\%
confidence level, instead of dividing them by 1.645 (Sect.~\ref{sec:data}) leaves $\kappa$ unchanged,
because the PHOBOS covariance is scaled as a whole; it raises $\sigma_\kappa$ to $\fvKappaErr$ through
the scale uncertainty ($\fvSigScale\%$ instead of $\sigScale\%$) and widens the uncertainties of the
window. The best $s$ then moves by at most $\fvShiftMax$, the 95\% intervals over the families widen
to \rng{\fvNFLoE}{\fvNFHiE}, \rng{\fvNFLoF}{\fvNFHiF} and \rng{\fvNFLoG}{\fvNFHiG},
$\Delta\chi^2(s{=}1)\le\fvDOneMax$, and the pull at 2.76 and 5.02~TeV falls to
\rng{\fvPullLoEF}{\fvPullHiEF}$\,\sigma_\kappa$: the conversion sets the precision of the test, not
its central values.

The calibration assumes that the estimator's bias, from the composition and $\langle m_T\rangle$ near
beam rapidity where neither is measured, is the same at RHIC and at the LHC. This is tested in both
generators with the participant energy as truth (Table~\ref{tab:closure}), with the kernel built as
for the data: at the LHC the generator's 200~GeV rapidity dependence is carried in $y/\ybeam$. The
nominal rule gives $R=\hijRcal$ (HIJING) and $\epoRcal$ (EPOS4) at 200~GeV within the measured range,
and \hijRlhc{} and \epoRlhc{} at the LHC over $|y|\le\ybeam$: the calibration
transfers with a factor \hijTransfer{} in HIJING and \epoTransfer{}
in EPOS4. With each generator's own rapidity dependence at the LHC the factors would be
\hijTransferOwn{} and \epoTransferOwn{}, so the difference
between the generators arises from how the rapidity dependence of the spectra changes with energy,
which current data do not constrain. The absolute $R$ depends on the generator; what is used here is the
transfer. With $\kappa\times\kTransfer$ (HIJING), $s={}$\rng{\sKLoE}{\sKHiE} at 2.76~TeV, \rng{\sKLoF}{\sKHiF}
at 5.02~TeV and \rng{\sKLoG}{\sKHiG} at 5.36~TeV, with $\Delta\chi^2(s{=}1)$ at most $\dOneKHiE$,
$\dOneKHiF$ and $\dOneKHiG$. With $\kappa\times\kTransferE$ (EPOS4) the values are $s={}$\rng{\sELoE}{\sEHiE},
\rng{\sELoF}{\sEHiF} and \rng{\sELoG}{\sEHiG}, with $\Delta\chi^2(s{=}1)$ at most $\dOneEHiE$, $\dOneEHiF$
and $\dOneEHiG$. The calibration with the STAR-only spectra, $\kappa_{\mathrm{STAR}}=\kappa\times
\kStarFactor$, gives nearly the same $s$ as the HIJING transfer (Table~\ref{tab:lf}). At 2.76~TeV, where
the data reach the edge of the window, the result depends only weakly on the calibration.

Under these assumptions limiting fragmentation is consistent with the data and the energy sum
rule at all three LHC energies: $s=1$ lies within the 95\% interval of every family for the nominal
rule and the nominal covariances, with or without either calibration transfer. Three caveats qualify
this. With other plausible correlations of the PHOBOS uncertainties, or another rapidity dependence of
the kernel, $s=1$ lies outside the 95\% interval of at least one family at each energy, although no
single setting excludes it at all three energies and the interval over the families excludes it only
\systUnionText{} (above). At 2.76 and 5.02~TeV
the LHC fits prefer \rng{\pullLoEF}{\pullHiEF}$\,\sigma_\kappa$ more estimated energy than the
calibration, and at 5.36~TeV \rng{\pullLoG}{\pullHiG}$\,\sigma_\kappa$ (the pull above); the transfers
indicated by the generators reduce this tension, to a pull of \rng{\pullKLo}{\pullKHi} (HIJING) and
\rng{\pullELo}{\pullEHi} (EPOS4) at 2.76 and 5.02~TeV. At 5.36~TeV a calibration lower by the same
factor as the HIJING transfer, $\kappa\times\kLowFactor$, would place
$s=1$ outside the 95\% interval for \nKlowOut{} of the three families
($\Delta\chi^2(s{=}1)\dOneKlowRel\dOneKlowHiG$). All statements are
conditional on the RHIC shape of the fragmentation region (Sect.~\ref{sec:discussion}); the data do
not show that the LHC follows it.

\section{Discussion}
\label{sec:discussion}

\paragraph{Relation to source models.} The result is consistent with the RDM and diffusion-model
analyses that attribute the dip mainly to the
Jacobian~\cite{Kellers:2019ezz,Kellers:2020xfa,Hoelck:2024cdl}. It also suggests why a calculation
that took the pion mass and an assumed $\langle p_T\rangle$ for its effective
Jacobian~\cite{Wolschin:2011za} reached a different conclusion from later ones whose effective
Jacobian was fixed by identified-particle data~\cite{Kellers:2019ezz,Kellers:2020xfa}: whether a dip-free $\dndy$ reproduces the $\eta$-dip depends on the
flatness of $\dndy$ near $y=0$ and on the full species and $p_T$ dependence of the Jacobian. The test
does not distinguish between production mechanisms that yield a dip-free $\dndy$ of the allowed
shapes, such as a dominant midrapidity source with fragmentation sources. In p+Pb collisions the RDM
finds the Jacobian insufficient to explain the dip~\cite{Schulz:2024imf}; the test does not address
asymmetric systems. The limiting-fragmentation results are consistent with the RDM expectation
that limiting fragmentation also holds at LHC energies~\cite{Kellers:2019ezz,Kellers:2020xfa}; here
the extrapolation is constrained by an energy sum rule calibrated on RHIC data, with the RHIC shape
of the fragmentation region imposed, rather than by a boundary condition.

\paragraph{What is not shown.} Broad depressions in $\dndy$ are only weakly constrained, because the
null families can absorb them, and the test is one-sided. A $\dndy$ with a distinct forward
component (HIJING) or a kernel that describes the forward region poorly (EPOS4) can produce a false
dip, as both generators do at 200~GeV (Sect.~\ref{sec:closure}). The split of the PHOBOS systematic
uncertainties is not known and is scanned; the dip-test conclusion holds over the whole scan.

\paragraph{Assumptions of the limiting-fragmentation test.} The test determines the normalization
of the LHC fragmentation region relative to RHIC; it assumes that the shape is the same as at RHIC. A change of shape at fixed
normalization would not be detected, and the data do not show that the LHC follows the RHIC shape.
Its result depends on the correlations of the PHOBOS uncertainties in the window, which are not
published (Sect.~\ref{sec:lf}).
It further assumes that the bias of the energy estimator transfers from RHIC to the LHC; the two
generators give factors of $\kTransfer$ and $\kTransferE$, and the spread between them is a model
uncertainty that current data cannot reduce. It also assumes the net-baryon profile at constant rapidity
loss and the kinematic-edge form of the fall-off. The absolute value of $f_E$ differs between the data
($\kappa$) and the generators ($R$ in Table~\ref{tab:closure}). The difference is not understood; much
of $f_E$ comes from beyond the measured BRAHMS range, where neither composition nor spectra are
measured (Sect.~\ref{sec:sumrule}), and only the transfer is taken from the generators.

\paragraph{Spectra at RHIC.} At 62.4 and 200~GeV the STAR spectra cover a narrow $p_T$ range, and
their L\'evy--Tsallis extrapolation inflates $\langle p_T\rangle$. At 200~GeV it gives a proton
$\langle p_T\rangle$ of $\mptPstar$~GeV, above the $\mptPlhc$~GeV of the 2.76~TeV spectra, whereas the
PHENIX spectra give $\mptPnom$~GeV; at 62.4~GeV the extended spectra give $\mptPnomB$~GeV instead of
$\mptPstarB$~GeV. EPOS4, which has radial flow, shows a bias of the same sign (protons
$+\epoPtStarD\%$ at 200~GeV, against $+\mptBiasData\%$ in the data), and HIJING a much smaller
one (Sect.~\ref{sec:closure}). The nominal
kernel therefore uses the PHENIX spectra at 200~GeV and the extended spectra at 62.4~GeV, with the
STAR-only spectra as a variant. The dip result does not depend on this choice. The energy
calibration does, with $\kappa_{\mathrm{STAR}}=\kappaStar\pm\kappaStarErr$ for the STAR-only spectra,
although the limiting-fragmentation ratio changes little (Table~\ref{tab:lf}).

\paragraph{Outlook.} The precision of the limiting-fragmentation test is limited by the RHIC
uncertainties in the test window and their unknown correlations, which a published breakdown of the
PHOBOS systematic uncertainties would constrain, and, at 5.02 and 5.36~TeV, by the coverage of the LHC data. Measurements of $\dndeta$ closer to beam rapidity at the LHC would test the extrapolation
directly, and identified spectra over a wider rapidity range at RHIC would constrain both the kernel
and the energy calibration.

\section{Summary}
\label{sec:summary}

Without an event generator, we have tested whether the midrapidity dip of $\dndeta$ in central
heavy-ion collisions requires a structure in $\dndy$, using a kernel built from measured identified
spectra, and have quantified the
sensitivity of the test. At all seven
energies from 19.6~GeV to 5.36~TeV the data do not require it: they are consistent with a dip-free
$\dndy$ folded with the $y\to\eta$ kinematics of the measured spectra. For dips with widths up to one unit of rapidity, the
depth giving $p<0.05$ in half of the pseudo-experiments is \rng{\dminLHClo}{\dminLHChi}\% at the LHC and
\rng{\dminRHIClo}{\dminRHIChi}\% at 62.4--200~GeV; broad depressions are only weakly constrained. Closure
tests in HIJING and EPOS4 probe the kernel and the procedure and show a failure mode, a false dip at
top RHIC energy that disappears when the most forward points are excluded from the fit.

For limiting fragmentation at the LHC the data alone leave the fragmentation region largely
unconstrained. An
energy sum rule calibrated on RHIC data, with the RHIC shape of the fragmentation region imposed, gives
best-fit LHC/RHIC ratios of \rng{\sLoE}{\sHiE} at 2.76~TeV, \rng{\sLoF}{\sHiF} at 5.02~TeV and
\rng{\sLoG}{\sHiG} at 5.36~TeV, with 95\% intervals that include unity at all three energies, with or
without the calibration transfers indicated by HIJING ($\kappa\times\kTransfer$) and EPOS4
($\kappa\times\kTransferE$). Under these assumptions $s=1$ is thus allowed,
although at 2.76 and 5.02~TeV the LHC fits prefer \rng{\pullLoEF}{\pullHiEF}$\,\sigma_\kappa$
more estimated energy than the calibration. The result depends on the unpublished correlations of the
PHOBOS uncertainties in the test window and on the kernel's rapidity dependence: over these variants
the best-fit ratios span \rng{\sSystLoE}{\sSystHiE}, \rng{\sSystLoF}{\sSystHiF} and \rng{\sSystLoG}{\sSystHiG},
and for some of them $s=1$ lies outside the 95\% interval of at least one family at each energy,
although no single setting excludes it at all three energies and the interval over the families
excludes it only \systUnionText. Limiting fragmentation is therefore neither established nor excluded
by these data. At 2.76~TeV the result is consistent with a direct comparison at the
most forward measured point; at 5.02 and 5.36~TeV it rests on the sum-rule extrapolation, which
measurements closer to beam rapidity would test.

\begin{acknowledgements}
The authors gratefully acknowledge Poojan Angiras, Manisha Rana and Sachin Rana for valuable
discussions and feedback. The computations used the CERN batch computing service.
\end{acknowledgements}

{\sloppy\paragraph{Data Availability Statement} This manuscript has associated data in a data repository.
[Authors' comment: All experimental data used in this study are public. The LHC pseudorapidity
densities, the identified-hadron spectra and the BRAHMS data are HEPData records; the PHOBOS tables
are from the PHOBOS publication data archive,
\url{https://www.sdcc.bnl.gov/phobos/Publications/Physics/}.]\par}

{\emergencystretch=2em\paragraph{Code Availability Statement} Code/\allowbreak soft\-ware will be made
available on reasonable request. [Authors' comment: The analysis code and the scripts that produce
every number, table and figure of this paper are available from the corresponding author on
reasonable request.]\par}

\paragraph{Competing interests} The authors declare no competing interests.

\appendix

\section{Documented LHC covariances}
\label{app:cov}

Each LHC point's systematic uncertainty $\sigma^{\mathrm{sys}}_j$ is split into a fully correlated
normalization part and a remainder. For ALICE at 2.76~TeV the normalization part is $\covNormE\%$ of
the value; of the remaining variance a share of $\covInE\%$ is fully correlated within $|\eta|<2$
and of $\covOutE\%$ beyond, and the rest is uncorrelated; the two $|\eta|$ regions are correlated
with each other only through the normalization. At 5.02~TeV the normalization part is
$\covNormF\%$, with a correlated share of $\covInF\%$ within $|\eta|<2$ and none beyond. For CMS at
5.36~TeV the normalization part is $\covNormG\%$, and the remaining variance is split into
uncorrelated, exponential-shape (correlation length $\covCmsEll$ unit of $\eta$) and fully correlated
parts in the proportions $\covCmsUnc$, $\covCmsShape$ and $\covCmsFull$. The variants change the
correlated share within $|\eta|<2$ to $\covVarLo\%$ and $\covVarHi\%$, the normalization at 2.76~TeV to $\covNormVarE\%$, and the CMS
correlation length to $\covEllLo$ and $\covEllHi$. The diagonal always equals the published total.

\section{Changes after the decision rule was fixed}
\label{app:amendments}

The method and decision rule were fixed in the authors' analysis notes before any $p$-value was
computed. They were not registered publicly, so this ordering rests on the authors' own records.
Each later change was fixed before the corresponding rerun:
(1) the covariance model was replaced by the three-way split of Sect.~\ref{sec:method}, because the
original form was nearly singular;
(2) the RHIC energies were added, with their fit ranges and variants;
(3) a fitter defect that could leave H0 in a local minimum was fixed; in the data it had inflated
$q$ (one spurious $p=0.04$ at 19.6~GeV), in the pseudo-experiments it had biased $p$ upward;
(4) digitized PHOBOS points were replaced by the official tables;
(5) the symmetrized LHC tables are used with each $|\eta|$ once, and the documented LHC covariances
became nominal;
(6) the PHENIX spectra at 200~GeV and the extended spectra at 62.4~GeV became nominal, after the
extrapolation of the narrow STAR $p_T$ ranges was found to inflate $\langle p_T\rangle$
(Sect.~\ref{sec:discussion}).
None of these changes altered a decision. Two further departures from the original plan were made
after the results: the injection grid was extended from three to \nWidthsWord{} widths, and the planned 95\%
confidence-level upper limits on $d$ from the profile likelihood are not quoted, because they are not
reliable at the boundary $d\ge0$; CL$_s$ limits from the injection pseudo-experiments are quoted
instead (Sect.~\ref{sec:results}). The mirror-symmetric covariance, the injection into a truth outside
the families and the rejection rate at the smallest depths (Sects.~\ref{sec:sensitivity}
and~\ref{sec:results}) were also added after the results. The generalized-Gaussian family was added as a robustness check after the
plan was fixed; it does not enter the decision. The edge-family check of Sect.~\ref{sec:results} and the
limiting-fragmentation analysis are not part of the original plan.

\bibliographystyle{spphys}
\bibliography{references}

\end{document}